\documentclass[%
twocolumn,
reprint,
superscriptaddress,
showpacs,preprintnumbers,
 amsmath,amssymb,
 aps,
]{revtex4-1}
\usepackage{placeins}
\usepackage{natbib}
\usepackage{graphicx}
\usepackage{dcolumn}
\usepackage{bm}
\usepackage[colorlinks]{hyperref}
\usepackage{graphicx}
\usepackage{amssymb}
\usepackage[latin2]{inputenc}
\usepackage[T1]{fontenc}
\usepackage{latexsym}
\usepackage{here}
\usepackage{mwe}
\usepackage{longtable}
\usepackage{textcomp}
\usepackage{amsmath}
\usepackage{amssymb}
\usepackage{epstopdf}
\usepackage{epsfig}
\usepackage{dcolumn}
\usepackage{bm}
\usepackage[pdf]{pstricks}
\begin{document}


\title{Detailed spectroscopy of doubly magic $^{132}$Sn}

\author{J.\,Benito}
\email{jabenito@ucm.es}
\affiliation{Grupo de F\'isica Nuclear \& IPARCOS, Universidad Complutense de Madrid, CEI Moncloa, E-28040 Madrid, Spain}
\author{L.M.\,Fraile}
\email{lmfraile@ucm.es}
\affiliation{Grupo de F\'isica Nuclear \& IPARCOS, Universidad Complutense de Madrid, CEI Moncloa, E-28040 Madrid, Spain}
\author{A.\,Korgul}
\affiliation{Faculty of Physics, University of Warsaw, PL 02-093 Warsaw, Poland}
\author{M.\,Piersa}
\affiliation{Faculty of Physics, University of Warsaw, PL 02-093 Warsaw, Poland}
\author{E.\,Adamska}
\affiliation{Faculty of Physics, University of Warsaw, PL 02-093 Warsaw, Poland}
\author{A.N.\,Andreyev}
\affiliation{Department of Physics, University of York, York, YO10 5DD, United Kingdom}
\affiliation{Advanced Science Research Center (ASRC), Japan Atomic Energy Agency, Tokai-mura, Japan}
\author{R.\,\'Alvarez-Rodr\'iguez}
\affiliation{Escuela T\'ecnica Superior de Arquitectura, Universidad Polit\'ecnica de Madrid, E-28040 Madrid, Spain}
\author{A.E.\,Barzakh}
\affiliation{Petersburg Nuclear Physics Institute, NRC Kurchatov Institute, 188300 Gatchina, Russia}
\author{G.\,Benzoni}
\affiliation{Istituto Nazionale di Fisica Nucleare, Sezione di Milano, I-20133 Milano, Italy}
\author{T.\,Berry}
\affiliation{Department of Physics, University of Surrey, Guildford GU2 7XH, United Kingdom}
\author{M.J.G.\,Borge}
\affiliation{CERN, CH-1211 Geneva 23, Switzerland}
\affiliation{Instituto de Estructura de la Materia, CSIC, E-28040 Madrid, Spain}
\author{M.\,Carmona}
\affiliation{Grupo de F\'isica Nuclear \& IPARCOS, Universidad Complutense de Madrid, CEI Moncloa, E-28040 Madrid, Spain}
\author{K.\,Chrysalidis}
\affiliation{CERN, CH-1211 Geneva 23, Switzerland}
\author{C.\,Costache}
\affiliation{``Horia Hulubei" National Institute of Physics and Nuclear Engineering, RO-077125 Bucharest, Romania}
\author{J.G.\,Cubiss}
\affiliation{CERN, CH-1211 Geneva 23, Switzerland}
\affiliation{Department of Physics, University of York, York, YO10 5DD, United Kingdom}
\author{T.\,Day\,Goodacre}
\affiliation{CERN, CH-1211 Geneva 23, Switzerland}
\affiliation{School of Physics and Astronomy, The University of Manchester, Manchester, United Kingdom}
\author{H.\,De\,Witte}
\affiliation{CERN, CH-1211 Geneva 23, Switzerland}
\affiliation{Instituut voor Kern- en Stralingsfysica, KU\,Leuven, B-3001 Leuven, Belgium}
\author{D.\,V.\,Fedorov}
\affiliation{Petersburg Nuclear Physics Institute, NRC Kurchatov Institute, 188300 Gatchina, Russia}
\author{V.\,N.\,Fedosseev}
\affiliation{CERN, CH-1211 Geneva 23, Switzerland}
\author{G.\,Fern\'andez-Mart\'inez}
\affiliation{Institut f\"ur Kernphysik, Technische Universit\"at Darmstadt, Germany}
\author{A.\,Fija\l{}kowska}
\affiliation{Faculty of Physics, University of Warsaw, PL 02-093 Warsaw, Poland}
\author{M.\,Fila}
\affiliation{Faculty of Physics, University of Warsaw, PL 02-093 Warsaw, Poland}
\author{H.\,Fynbo}
\affiliation{Department of Physics and Astronomy, Aarhus University, DK-8000 Aarhus C, Denmark}
\author{D.\,Galaviz}
\affiliation{LIP, and Faculty of Sciences, University of Lisbon, 1000-149 Lisbon, Portugal}
\author{P.\,Galve}
\affiliation{Grupo de F\'isica Nuclear \& IPARCOS, Universidad Complutense de Madrid, CEI Moncloa, E-28040 Madrid, Spain}
\author{M.\,Garc\'ia-D\'iez}
\affiliation{Grupo de F\'isica Nuclear \& IPARCOS, Universidad Complutense de Madrid, CEI Moncloa, E-28040 Madrid, Spain}
\author{P.T.\,Greenlees}
\affiliation{University of Jyv\"askyl\"a, Department of Physics, P.O. Box 35, FI-40014 Jyv\"askyl\"a, Finland}
\affiliation{Helsinki Institute of Physics, University of Helsinki, FIN-00014 Helsinki, Finland}
\author{R.\,Grzywacz}
\affiliation{Department of Physics and Astronomy, University of Tennessee, Knoxville, Tennessee 37996, USA}
\affiliation{Physics Division, Oak Ridge National Laboratory, Oak Ridge, Tennessee 37831, USA}
\author{L.J.\,Harkness-Brennan}
\affiliation{Department of Physics, Oliver Lodge Laboratory, University of Liverpool, Liverpool L69 7ZE, United Kingdom}
\author{C.\,Henrich}
\affiliation{Institut f\"ur Kernphysik, Technische Universit\"at Darmstadt, 64289 Darmstadt, Germany}
\author{M.\,Huyse}
\affiliation{Instituut voor Kern- en Stralingsfysica, KU\,Leuven, B-3001 Leuven, Belgium}
\author{P.\,Ib\'a\~nez}
\affiliation{Grupo de F\'isica Nuclear \& IPARCOS, Universidad Complutense de Madrid, CEI Moncloa, E-28040 Madrid, Spain}
\author{A.\,Illana}
\email{Present address: University of Jyv\"askyl\"a, Department of Physics, P.O. Box 35, FI-40014 Jyv\"askyl\"a, Finland}
\affiliation{Instituut voor Kern- en Stralingsfysica, KU\,Leuven, B-3001 Leuven, Belgium}
\affiliation{Instituto Nazionale di Fisica Nucleare, Laboratori Nazionali di Legnaro, I-35020 Legnaro, Italy}
\author{Z.\,Janas}
\affiliation{Faculty of Physics, University of Warsaw, PL 02-093 Warsaw, Poland}
\author{J.\,Jolie}
\affiliation{Institut f\"ur Kernphysik, Universit\"at zu K\"oln, D-50937 K\"oln, Germany}
\author{D.S.\,Judson}
\affiliation{Department of Physics, Oliver Lodge Laboratory, University of Liverpool, Liverpool L69 7ZE, United Kingdom}
\author{V.\,Karayonchev}
\affiliation{Institut f\"ur Kernphysik, Universit\"at zu K\"oln, D-50937 K\"oln, Germany}
\author{M.\,Kici\'nska-Habior}
\affiliation{Faculty of Physics, University of Warsaw, PL 02-093 Warsaw, Poland}
\author{J.\,Konki}
\email{Present address: CERN, CH-1211 Geneva 23, Switzerland }
\affiliation{University of Jyv\"askyl\"a, Department of Physics, P.O. Box 35, FI-40014 Jyv\"askyl\"a, Finland}
\affiliation{Helsinki Institute of Physics, University of Helsinki, FIN-00014 Helsinki, Finland}
\author{J.\,Kurcewicz}
\affiliation{CERN, CH-1211 Geneva 23, Switzerland}
\author{I.\,Lazarus}
\affiliation{STFC Daresbury, Daresbury, Warrington WA4 4AD, United Kingdom}
\author{R.\,Lic\u{a}}
\affiliation{CERN, CH-1211 Geneva 23, Switzerland}
\affiliation{``Horia Hulubei" National Institute of Physics and Nuclear Engineering, RO-077125 Bucharest, Romania}
\author{A.\,L\'opez-Montes}
\affiliation{Grupo de F\'isica Nuclear \& IPARCOS, Universidad Complutense de Madrid, CEI Moncloa, E-28040 Madrid, Spain}
\author{M.\,Lund}
\affiliation{Department of Physics and Astronomy, Aarhus University, DK-8000 Aarhus C, Denmark}
\author{H.\,Mach}
\email{Deceased.}
\affiliation{National Centre for Nuclear Research, PL 02-093 Warsaw, Poland}
\author{M.\,Madurga}
\affiliation{CERN, CH-1211 Geneva 23, Switzerland}
\affiliation{Department of Physics and Astronomy, University of Tennessee, Knoxville, Tennessee 37996, USA}
\author{I.\,Marroqu\'in}
\affiliation{Instituto de Estructura de la Materia, CSIC, E-28040 Madrid, Spain}
\author{B.\,Marsh}
\affiliation{CERN, CH-1211 Geneva 23, Switzerland}
\author{M.C.\,Mart\'inez}
\affiliation{Grupo de F\'isica Nuclear \& IPARCOS, Universidad Complutense de Madrid, CEI Moncloa, E-28040 Madrid, Spain}
\author{C.\,Mazzocchi}
\affiliation{Faculty of Physics, University of Warsaw, PL 02-093 Warsaw, Poland}
\author{N.\,M\u{a}rginean}
\affiliation{``Horia Hulubei" National Institute of Physics and Nuclear Engineering, RO-077125 Bucharest, Romania}
\author{R.\,M\u{a}rginean}
\affiliation{``Horia Hulubei" National Institute of Physics and Nuclear Engineering, RO-077125 Bucharest, Romania}
\author{K.\,Miernik}
\affiliation{Faculty of Physics, University of Warsaw, PL 02-093 Warsaw, Poland}
\author{C.\,Mihai}
\affiliation{``Horia Hulubei" National Institute of Physics and Nuclear Engineering, RO-077125 Bucharest, Romania}
\author{R.E.\,Mihai}
\affiliation{``Horia Hulubei" National Institute of Physics and Nuclear Engineering, RO-077125 Bucharest, Romania}
\author{E.\,N\'acher}
\affiliation{Instituto de F\'isica Corpuscular, CSIC - Universidad de Valencia, E-46071 Valencia, Spain}
\author{A.\,Negret}
\affiliation{``Horia Hulubei" National Institute of Physics and Nuclear Engineering, RO-077125 Bucharest, Romania}
\author{B.\,Olaizola}
\affiliation{TRIUMF, 4004 Wesbrook Mall, Vancouver, British Columbia, V6T 2A3 Canada}
\author{R.D.\,Page}
\affiliation{Department of Physics, Oliver Lodge Laboratory, University of Liverpool, Liverpool L69 7ZE, United Kingdom}
\author{S.V.\,Paulauskas}
\affiliation{Department of Physics and Astronomy, University of Tennessee, Knoxville, Tennessee 37996, USA}
\author{S.\,Pascu}
\affiliation{``Horia Hulubei" National Institute of Physics and Nuclear Engineering, RO-077125 Bucharest, Romania}
\author{A.\,Perea}
\affiliation{Instituto de Estructura de la Materia, CSIC, E-28040 Madrid, Spain}
\author{V.\,Pucknell}
\affiliation{STFC Daresbury, Daresbury, Warrington WA4 4AD, United Kingdom}
\author{P.\,Rahkila}
\affiliation{University of Jyv\"askyl\"a, Department of Physics, P.O. Box 35, FI-40014 Jyv\"askyl\"a, Finland}
\affiliation{Helsinki Institute of Physics, University of Helsinki, FIN-00014 Helsinki, Finland}
\author{C.\,Raison}
\affiliation{Department of Physics, University of York, York, YO10 5DD, United Kingdom}
\author{E.\,Rapisarda}
\affiliation{CERN, CH-1211 Geneva 23, Switzerland}
\author{J.-M.\,R\'egis}
\affiliation{Institut f\"ur Kernphysik, Universit\"at zu K\"oln, D-50937 K\"oln, Germany}
\author{K.\,Rezynkina}
\affiliation{Instituut voor Kern- en Stralingsfysica, KU\,Leuven, B-3001 Leuven, Belgium}
\author{F.\,Rotaru}
\affiliation{``Horia Hulubei" National Institute of Physics and Nuclear Engineering, RO-077125 Bucharest, Romania}
\author{S.\,Rothe}
\affiliation{CERN, CH-1211 Geneva 23, Switzerland}
\author{D.\,S\'anchez-Parcerisa}
\affiliation{Grupo de F\'isica Nuclear \& IPARCOS, Universidad Complutense de Madrid, CEI Moncloa, E-28040 Madrid, Spain}
\affiliation{Sedecal Molecular Imaging, E-28110 Algete (Madrid), Spain}
\author{V.\,S\'anchez-Tembleque}
\affiliation{Grupo de F\'isica Nuclear \& IPARCOS, Universidad Complutense de Madrid, CEI Moncloa, E-28040 Madrid, Spain}
\author{K.\,Schomacker}
\affiliation{Institut f\"ur Kernphysik, Universit\"at zu K\"oln, D-50937 K\"oln, Germany}
\author{G.\,S.\,Simpson}
\affiliation{LPSC, IN2P3-CNRS/Universit\'e Grenoble Alpes, Grenoble Cedex F-38026, France}
\author{Ch.\,Sotty}
\affiliation{Instituut voor Kern- en Stralingsfysica, KU\,Leuven, B-3001 Leuven, Belgium}
\affiliation{``Horia Hulubei" National Institute of Physics and Nuclear Engineering, RO-077125 Bucharest, Romania}
\author{L.\,Stan}
\affiliation{``Horia Hulubei" National Institute of Physics and Nuclear Engineering, RO-077125 Bucharest, Romania}
\author{M.\,St\u{a}noiu}
\affiliation{``Horia Hulubei" National Institute of Physics and Nuclear Engineering, RO-077125 Bucharest, Romania}
\author{M.\,Stryjczyk}
\affiliation{Faculty of Physics, University of Warsaw, PL 02-093 Warsaw, Poland}
\affiliation{Instituut voor Kern- en Stralingsfysica, KU\,Leuven, B-3001 Leuven, Belgium}
\author{O.\,Tengblad}
\affiliation{Instituto de Estructura de la Materia, CSIC, E-28040 Madrid, Spain}
\author{A.\,Turturica}
\affiliation{``Horia Hulubei" National Institute of Physics and Nuclear Engineering, RO-077125 Bucharest, Romania}
\author{J.M.\,Ud\'ias}
\affiliation{Grupo de F\'isica Nuclear \& IPARCOS, Universidad Complutense de Madrid, CEI Moncloa, E-28040 Madrid, Spain}
\author{P.\,Van\,Duppen}
\affiliation{Instituut voor Kern- en Stralingsfysica, KU\,Leuven, B-3001 Leuven, Belgium}
\author{V.\,Vedia}
\affiliation{Grupo de F\'isica Nuclear \& IPARCOS, Universidad Complutense de Madrid, CEI Moncloa, E-28040 Madrid, Spain}
\author{A.\,Villa-Abaunza}
\affiliation{Grupo de F\'isica Nuclear \& IPARCOS, Universidad Complutense de Madrid, CEI Moncloa, E-28040 Madrid, Spain}
\author{S.\,Vi\~nals}
\affiliation{Instituto de Estructura de la Materia, CSIC, E-28040 Madrid, Spain}
\author{W.B.\,Walters}
\affiliation{Department of Chemistry, University of Maryland, Maryland 20742, USA}
\author{R.\,Wadsworth}
\affiliation{Department of Physics, University of York, York, YO10 5DD, United Kingdom}
\author{N.\,Warr}
\affiliation{Institut f\"ur Kernphysik, Universit\"at zu K\"oln, D-50937 K\"oln, Germany}

\collaboration{IDS collaboration}
\date{\today}
\begin{abstract}
The structure of the doubly magic $^{132}_{50}$Sn$_{82}$ has been investigated at the ISOLDE facility at CERN, populated both by the $\beta^-$decay of $^{132}$In and $\beta^-$-delayed neutron emission of $^{133}$In. The level scheme of $^{132}$Sn is greatly expanded with the addition of 68 $\gamma$-transitions and 17 levels observed for the first time in the $\beta$ decay. The information on the excited structure is completed by new $\gamma$-transitions and states populated in the $\beta$-n decay of $^{133}$In. Improved delayed neutron emission probabilities are obtained both for $^{132}$In and $^{133}$In. Level lifetimes are measured via the Advanced Time-Delayed $\beta\gamma\gamma$(t) fast-timing method. An interpretation of the level structure is given based on the experimental findings and the particle-hole configurations arising from core excitations both from the \textit{N} = 82 and \textit{Z} = 50 shells, leading to positive and negative parity particle-hole multiplets. The experimental information provides new data to challenge the theoretical description of $^{132}$Sn.              

\end{abstract}
\maketitle

\section{\label{sec:level1}Introduction}\label{Intro}

The \textsuperscript{132}Sn nucleus is one of the bastions of our understanding of nuclear structure in the framework of the nuclear shell model. With 50 protons and 82 neutrons it is one of the most exotic doubly-magic nuclei within reach of current experimental facilities. One of the signatures of its doubly-magic nature is the high lying first-excited state at 4041.6~keV \cite{Kerek1973,Bjornstad1986,NDS132}. This value, once scaled by a factor of \textit{A}$^{1/3}$ to account for the size of the nucleus, is large in comparison to other doubly magic nuclei such as \textsuperscript{208}Pb and \textsuperscript{16}O, which points to a strong double shell closure. The \textsuperscript{132}Sn doubly magic structure is also manifested by the almost pure single-particle nature of the levels in \textsuperscript{133}Sn. This nature was probed by a transfer reaction in inverse kinematics \cite{Jones2010} yielding very large spectroscopic factors. 

The region of the nuclear chart around $^{132}$Sn plays an important role in the astrophysical rapid neutron-capture process (r-process), which impacts elemental abundances in the solar system. Recently, the identification of the nucleosynthesis site for neutron-rich nuclei around \textit{N}~=~82 has been reported \cite{Barnes2016,Cowperth2017} associated to a "kilonova" \cite{Metzger2010}, observed by multimessenger astronomy. The robustness of the \textit{N}~=~82 neutron shell is one of the important parameters when modeling r-process nucleosynthesis and describing light curves arising from compact object mergers. The shell structure in this region is also necessary to understand the role of fission in the r-process \cite{Martinez2007}.

\begin{figure}[ht!]
\includegraphics[width=\columnwidth]{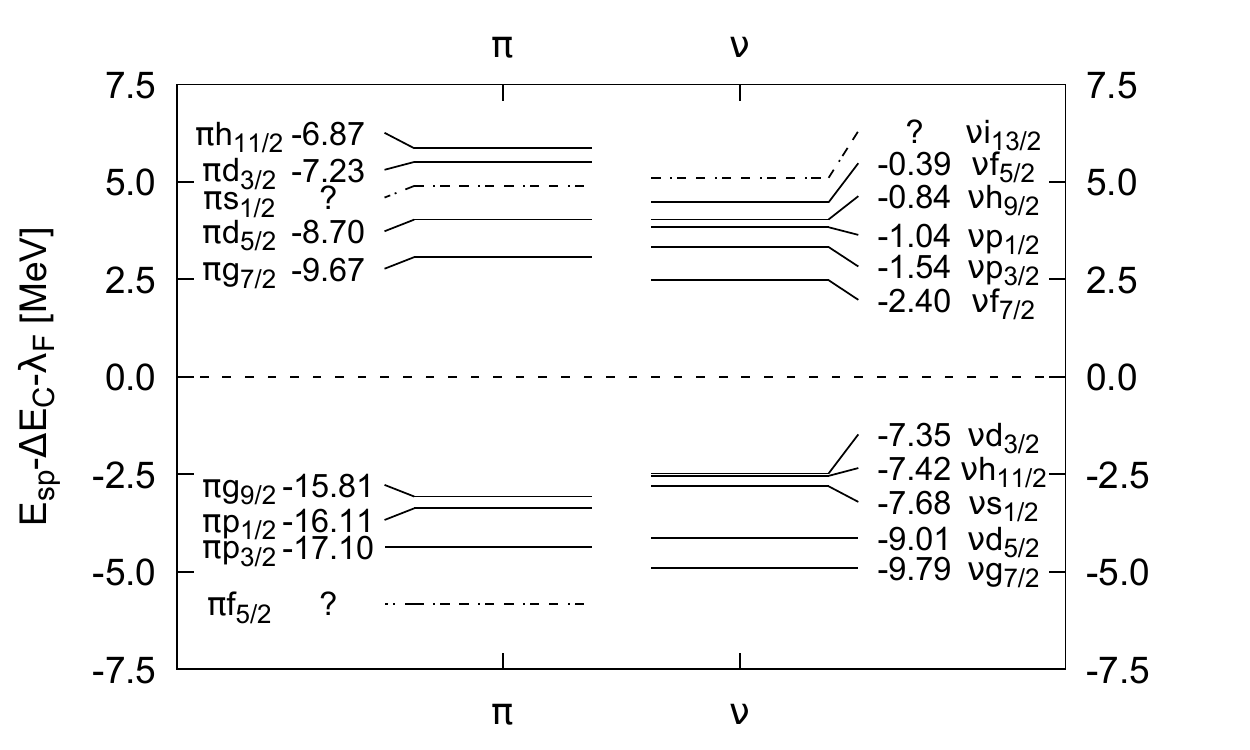}
\caption[]{\label{fig:spstates} Experimental single-particle and single-hole energies for neutrons ($\nu$) and protons ($\pi$) in the doubly-magic nucleus \textsuperscript{132}Sn. Following \cite{Grawe2007} the energy origin is set in the center of the shell gap ($\lambda_F$) in order to remove Coulomb energy differences $\Delta E_C$. Binding energies are taken from \cite{Wang2017}. The absolute single-particle energies are given in MeV.}
\end{figure}

From the point of view on nuclear structure the nuclei with a valence particle or hole around $^{132}$Sn are relevant for investigating single-particle states and transition probabilities. They provide observables that are the main ingredients in state-of-the-art large scale shell-model calculations to understand the nuclear structure in the region. 
The single-particle states in the region are represented in Figure \ref{fig:spstates} following the procedure described in \cite{Grawe2007}, 
and employing binding energies from \cite{Wang2017} and the excitation spectra from \textsuperscript{131}Sn, \textsuperscript{133}Sn, 
\textsuperscript{133}Sb and \textsuperscript{131}In. 
In particular, the neutron single-particle orbits above the \textit{N}~=~82 shell gap have been taken from low-lying states in \textsuperscript{133}Sn: the $\nu f_{7/2}$ ground state and the $\nu p_{3/2}$, $\nu p_{1/2}$, $\nu h_{9/2}$ and $\nu f_{5/2}$ states at 854, 1367, 1561 and 2005 keV excitation energy \cite{Hoff1996,Hoff2000,Jones2010, Jones2011,Allmond2014, Vaquero2017,Piersa2019}. The  $\nu i_{13/2}$ single-particle state has not been experimentally identified to date \cite{Korgul2015}.
These single particle states are not only relevant for \textsuperscript{132}Sn but also for neutron-rich nuclei in its vicinity.

Experimental data on \textsuperscript{132}Sn is essential for the shell-model description of the exotic nuclear region around \textit{N}~=~82 and to provide insight into particle-hole couplings for both protons and neutrons. Most of the excited levels in $^{132}$Sn correspond to particle-hole (p-h) configurations where a proton or a neutron is promoted across the closed shell. The coupling of single particle and single hole configurations (Figure \ref{fig:spstates}) leads to multiplets of excited states with an expected low admixture of other configurations. The particle-hole residual interaction makes the level energies within a multiplet non-degenerate. The identification of these multiplets provides information on the nuclear two-body matrix elements to first order. Experimentally measured transition rates between states within a multiplet, and also between states of different multiplets, give important knowledge of the underlying single particle structure. 

The investigation of the doubly-magic $^{132}$Sn is thus essential for theoretical models aiming at the understanding neutron-rich nuclei in the region. The development of these models is furthermore needed for the description of r-process nuclei that are at present experimentally out of reach.

The excited structure of $^{132}$Sn has been experimentally investigated since the 1970s. Beta decay experiments were carried out at OSIRIS \cite{Kerek1973,Fogelberg1994,Fogelberg1995,Mach1995a} and ISOLDE \cite{Bjornstad1986,BJORNSTAD198035}, as well as fission experiments performed at the JOSEF facility \cite{Kawade1982,DEHESA1978309} and the Argonne National Laboratory \cite{Bhattacharyya2001}. 

The most complete $\beta$-decay experiment was performed by Fogelberg \textit{et al.} \cite{Fogelberg1994,Fogelberg1995,Mach1995a} at the OSIRIS facility in the 1990s, where the level scheme of $^{132}$Sn was expanded to 21 excited levels, including negative and positive parity states up to the neutron separation energy. Proton particle-hole states were identified for the first time, and the 4352-keV \textit{J}$^\pi$~=~3$^-$ state was confirmed to have an octupole vibrational character. Lifetime measurements of the excited states down to the ps range were performed, and spin and parities assignments were made for the levels below 5 MeV.

Despite all the detailed studies on \textsuperscript{132}Sn attained through the $\beta$-decay and fission experiments, many of the expected particle-hole multiplet states remain without experimental identification. Beta decay is the ideal tool to investigate the excited structure of \textsuperscript{132}Sn, both directly from \textsuperscript{132}In (7$^-$) g.s. and via beta-delayed neutron emission from the \textsuperscript{133}In (9/2$^+$) g.s. and the (1/2$^-$) 330-keV beta-decaying isomer. This is due to the large energy window available for the decay, of $Q_{\beta}$($^{132}$In)~=~14140(60) keV and $Q_{\beta n}$($^{133}$In)~=~11010(200) keV (from systematics) \cite{Wang2017}, respectively, and due to the high spin of the parent nuclei that makes it possible to feed many states in \textsuperscript{132}Sn. 

In this work we focus on the investigation of the excited structure of \textsuperscript{132}Sn populated in $\beta$ and $\beta$-n decay. Taking advantage of the enhanced yield and selectivity achieved at the ISOLDE facility at CERN some of the missing particle-hole multiplet states have been identified. In addition, lifetimes of excited states in \textsuperscript{132}Sn have been measured using fast-timing techniques. The results from the $\beta$ decay study of the $^{133}$In isomers have been partially covered in \cite{Piersa2019}. Details on the experimental method used in the present work are provided in Section \ref{sec:experiment}. The experimental results are presented in Section \ref{sec:results} and discussed in Section \ref{sec:discussion}. Conclusions are drawn in Section \ref{sec:conclusions}.

\section{Experimental details}
\label{sec:experiment}
The experiment was carried out at the ISOLDE facility at CERN. It was performed in two separate data-taking campaigns in 2016 and 2018, where the excited structure in $^{132}$Sn was populated in the $\beta$ decay of $^{132}$In and in the $\beta$-n decay of the $^{133}$In (9/2$^{+}$) and (1/2$^{-}$) states. 
The $^{132}$In and $^{133}$In isotopes were produced by the bombardment of a UC$_{x}$ target equipped with a neutron converter by 1.4-GeV protons from the CERN PS-Booster (PSB). The indium ions thermally diffused out of the target and were ionized using the ISOLDE resonance ionization laser ion source (RILIS) \cite{Fedosseev2017}. The use of RILIS granted isomeric selectivity, by taking advantage of the difference in the hyperfine splitting of the isomer and ground state. More details on the isomer selection can be found in \cite{Piersa:2018kgo}. Following the ionization, indium ions were extracted and accelerated by a 40-kV potential difference, mass analyzed \cite{Catherall2017} and implanted on an aluminized mylar tape located at the center of our detector setup at the ISOLDE Decay Station (IDS) \cite{IDS}. The ions reached the IDS following the time structure of the PSB supercycle whereby proton pulses were grouped into sets of 34 or 35 pulses, out of which around half of them are delivered to the ISOLDE target-unit separated in time by multiples of 1.2 s. The beam was collected on the tape for a fixed time varying from 200 to 400 ms after the impact of each proton pulse. Once every supercycle the tape was moved in order to reduce the activity of long-lived daughter nuclides. The average beam intensity at the experimental station was of the order of 4$\cdot$10$^4$ and 2$\cdot$10$^3$ ions per second for $^{132}$In and $^{133}$In, respectively. Data 
were collected for 20 hours for each mass.

The IDS setup consists of a set of detectors aimed at measuring the $\beta$ and $\gamma$ radiation emitted after the $\beta$-decay of the implanted isotopes. They are arranged in close geometry surrounding the implantation point. The setup can be divided in two branches. The first branch is composed of four clover-type HPGe detectors for $\gamma$-ray spectroscopy, with a combined full-energy peak efficiency of 4$\%$ at 1173 keV. The second branch is aimed at lifetime measurement of excited states using the Advanced Time-Delayed $\beta\gamma\gamma(t)$ (fast timing) technique \cite{Mach1989,Moszynski1989,Fraile2017}. It consists of two LaBr$_3$(Ce) crystals with the shape of truncated cones \cite{Vedia2017} coupled to fast photomultiplier tubes (PMTs) \cite{fraile2013fast}, with 1$\%$ of total efficiency at 1 MeV each, and an ultrafast 3-mm thick NE111A plastic scintillator used as a $\beta$ detector, with a $\sim$20$\%$ efficiency. The energy and fast-timing signals are taken from the PMT dynode and anode outputs, respectively. The timing signals are processed by analog constant fraction discriminators and introduced in pairs in time-to-amplitude converter (TAC) modules that provide the time difference between them. In this experiment, time differences between the $\beta$ and the two LaBr$_3$(Ce) detectors as well as between the two LaBr$_3$(Ce) detectors were recorded. More details on the setup are provided in \cite{Fraile2017,Lica2017}.

All the signals from both branches were read and digitized by the Nutaq digital data acquisition system \cite{nutaq}. Logic signals, such as the time of arrival of the proton pulse on target and the tape movement, were also digitized. Data were collected in a triggerless mode. Events were built during the offline analysis where they were sorted in coincidence windows, and correlated with the proton arrival time. At this stage, the energy calibration for each detector was applied, as well as add-back corrections for the clover detectors.
Due to the large energy of the $^{132}$Sn $\gamma$ rays, a precise efficiency calibration is needed in a wide energy range. Therefore, $^{152}$Eu, $^{138}$Cs, $^{140}$Ba and $^{133}$Ba radioactive sources were used to build the energy calibration. The calibration was extended up to 7.6 MeV by including high-energy $\gamma$ rays originating from the capture of thermal neutrons produced at the ISOLDE target station, mainly in iron (IDS frame) and germanium (HPGe detectors).

For the timing measurements, the calibration of the LaBr$_3$(Ce) time response for full-energy peaks (FEP) as a function of energy, the FEP walk curve, is required. It was built for each LaBr$_3$(Ce) detector using $\beta\gamma$(t) and $\gamma\gamma$(t) coincidences with $^{140}$Ba/$^{140}$La and $^{152}$Eu $\gamma$-ray sources, $^{138}$Cs and  $^{88}$Rb on-line sources and by including several transitions in $^{132}$Sb as an internal calibration source. In this way, we obtained FEP time response curves for each LaBr$_3$(Ce) detector in the energy range 100 keV to 2.6 MeV, with an average one-sigma error of 3 ps. A similar procedure was implemented to build the Compton walk curve in order to take care of the corrections due to Compton events.

\section{Experimental results}
\label{sec:results}
The excited structure of $^{132}$Sn was populated through the $\beta$ decay of $^{132}$In (7$^{-}$), and from the $\beta$-n decay of the $^{133}$In (9/2$^{+}$) g.s. and the (1/2$^{-}$) isomer. The large differences in the spin and parity of the parent nuclei result in distinct feeding patterns for each decay, which provides information on the spin and parity of the levels fed.

Owing to the large difference in the $\beta$-decay half-lives of $^{132}$In and $^{132}$Sn, of 200(2)~ms and 39.7(8)~s \cite{NDS132}, respectively, the time distribution relative to the arrival of the proton on target makes it possible to identify whether a $\gamma$ ray has been emitted during the $\beta$ decay of $^{132}$In or from the daughters. A similar situation occurs for the $\beta$ decay of $^{133}$In.
Apart from the time distribution, the identification of the $\gamma$-rays belonging to $^{132}$Sn is based on $\gamma$-$\gamma$ coincidences with previously-known transitions. 

\subsection{Beta decay of $^{132}$In}

The Q$_{\beta}$ in $^{132}$In is 14140(60) keV, while the neutron separation energy in $^{132}$Sn is 7353(4) keV \cite{Wang2017}. Hence, the feeding of excited states up to $\approx$ 7 MeV is possible in this decay and high energy $\gamma$-rays may be observed. 
The excited structure populated in the $^{132}$In $\beta$ decay is very complex. The high spin of the parent, (7$^{-}$) \cite{Jungclaus2016}, favors the population of high spin (6-8) excited states in the energy range from 4 to 7 MeV. Those levels can only de-excite to the ground state by means of $\gamma$-ray cascades of 3 or more transitions. As discussed before, due to the doubly magic nature of $^{132}$Sn, the first excited state appears at a very high energy, 4041 keV.Therefore, it is not expected to find new levels in this decay that can de-excite directly to the g.s. with an energy below 4 MeV.

\begin{figure}[ht!]
\includegraphics[width=\columnwidth]{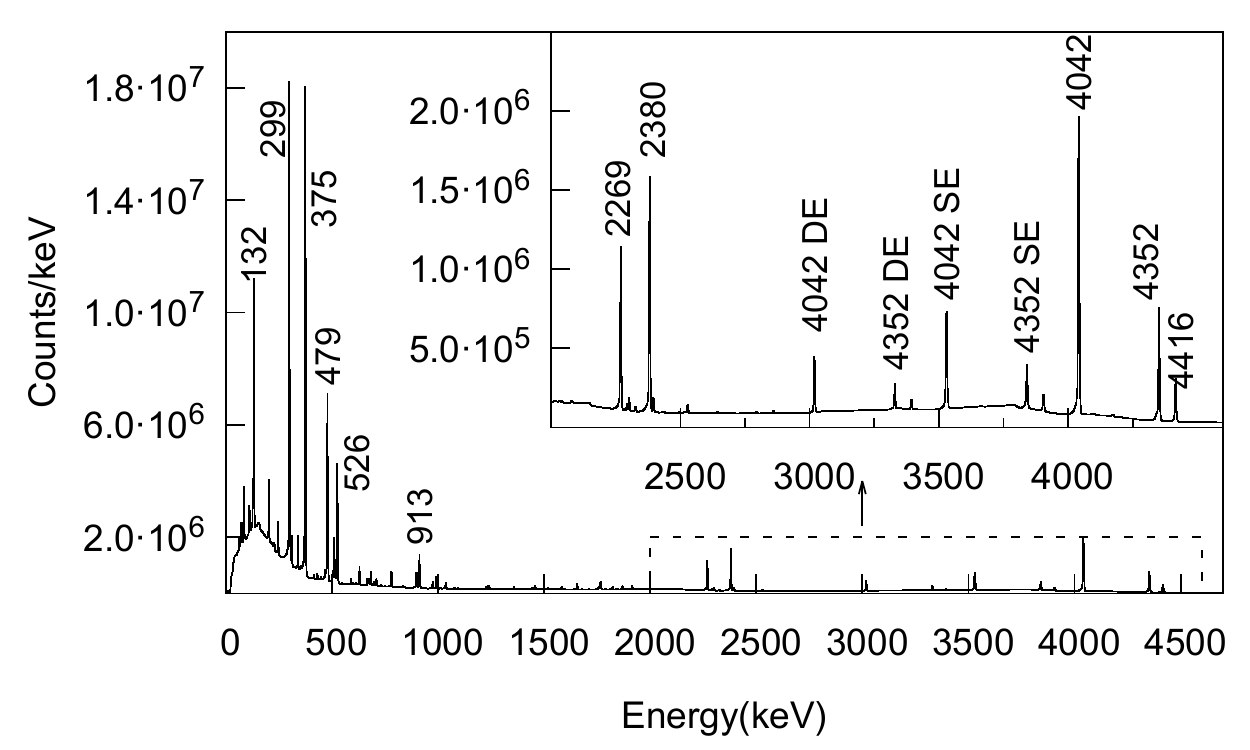}
\caption[]{\label{fig:epsart}  Singles $\gamma$-ray energy spectrum recorded by the HPGe detectors following the $^{132}$In decay.  This histogram was built using the events measured during the 30-530~ms time interval after proton pulse. The strongest peaks observed in the spectra are labeled with their energies in keV. The SE and DE labels indicates single escape and double escape peaks. The inset zooms in the 2400-4450 keV region.
}
\end{figure}

The energy spectra recorded by the HPGe clover detectors, setting a time window of 30-530 ms after proton impact are depicted in Figure \ref{fig:epsart}. This condition was imposed in order to reduce the contribution of the different contaminants. The contribution of neutron-induced background coming from the target is suppressed by removing the first 30 ms of the time window. An upper limit of the time window at 530 ms was chosen to reduce the contribution of the long-lived daughters, while keeping most of the statistics. 

\subsubsection{Half-life of $^{132}$In ground state}

The half-life of $^{132}$In was measured by fitting the time distribution of the 10 strongest $\gamma$ rays (labeled in Fig. \ref{fig:epsart} except for the 4416 keV). Since the activity at the experimental station is pulsed by the proton beam structure and the release from the target, the time distribution is fitted to an exponential decay function with a constant background after the end of the implantation. The background contribution was estimated by analyzing the time range from 2400 to 3600 ms after proton impact. 
\begin{figure}[ht!]
\includegraphics[width=\columnwidth]{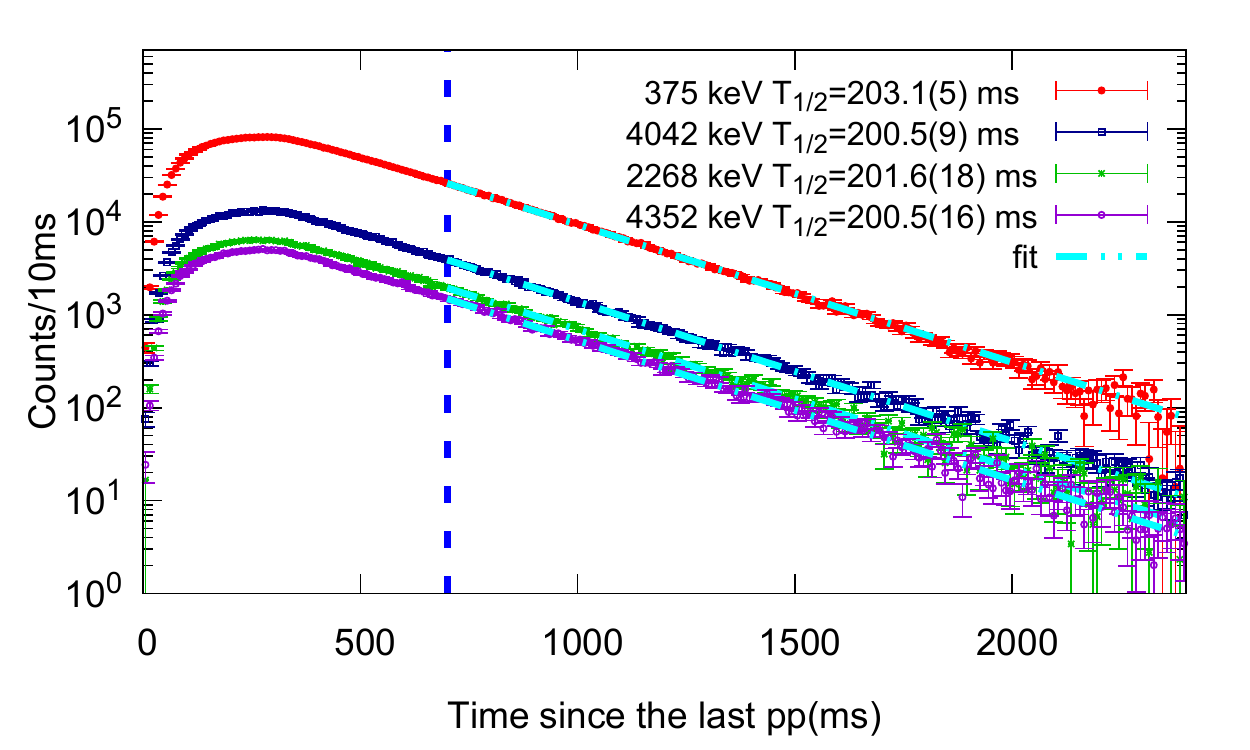}
\caption[]{\label{fig:lifeIn} Decay curves of the 375-, 4042-, 2268- and 4352-keV $\gamma$-rays recorded in singles in the HPGe detectors. The region considered for the fit goes from 700 ms, marked with a dark-blue dashed line, up to 2400 ms. This region has been adjusted in order to minimize dead-time effects.}
\end{figure} 
Due to the high count rates, dead time effects are sizable, mainly during implantation, but also at the beginning of the decay. In each HPGe crystal an average count rate of 3$\cdot$10$^3$ counts per seconds (cps) was observed, however, during the implantation time the count rate could rise up to 1.5$\cdot$10$^4$ cps. To account for this effect the beginning of the fit range was shifted by a few half-lives towards higher times and a $\chi^2$ fit test was performed to verify that the expected exponential decay behavior was recovered. The lifetime measurement was performed independently for each of the two data sets of $^{132}$In decay measured in the two experimental campaigns. The analysis made use of single events recorded in the HPGe detectors. The contribution of the Compton background under full-energy peaks was subtracted. This investigation furnishes 20 independent values for the $^{132}$In, 10 from each data set, which are all statistically compatible with each other. The final value of the $^{132}$In half-life was adopted as the weighted average of these measurements yielding T$_{1/2}$=202.2(2)~ms. The statistical uncertainty of the weighted average is calculated and increased by multiplying by the $\chi^2$ obtained. No systematic error is included. This half-lifes is in agreement with, but more precise than, the value reported in the latest evaluation, T$_{1/2}$=200(2)~ms \cite{NDS132}. In Figure \ref{fig:lifeIn} the decay curves for 4 of the $\gamma$ rays under consideration, from the 2016 data set, are shown.

\subsubsection{Identification of new $\gamma$-rays in $^{132}$Sn}

\begin{figure*}[hbt!]
\includegraphics[width=\textwidth]{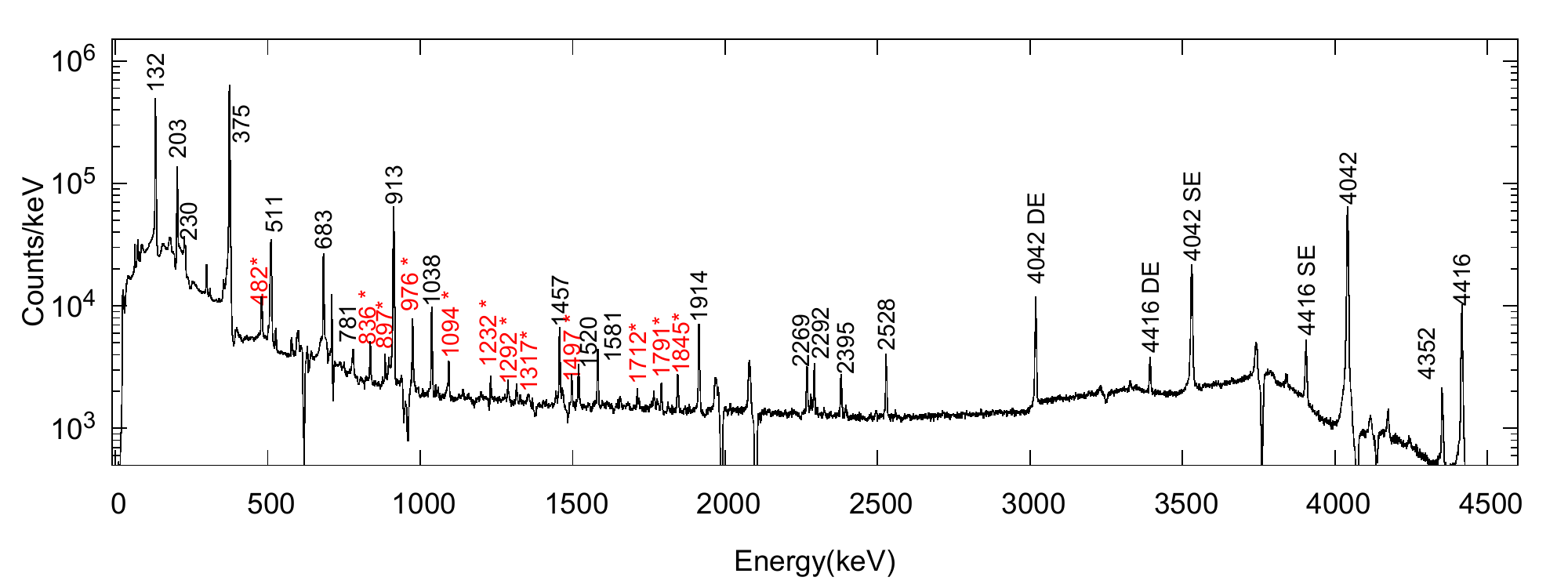}
\caption{\label{fig:299_coin} Compton-subtracted $\gamma$-$\gamma$ energy spectrum gated on the 6$^{+}$ $\rightarrow$ 4$^{+}$ 299.3~keV $\gamma$ transition in $^{132}$Sn. 
The previously-known $\gamma$ rays are labeled with their energies in black. New transitions identified in this work are labeled  in red with an asterisk. The negative peaks in the spectrum arise from background subtraction, due to Compton scattering between two HPGe clover detectors.}
\end{figure*}

The analysis of $\gamma$-$\gamma$ coincidences was done using the full statistics independently for the two data sets from each campaign. The assignments were cross-checked by requiring coincidences with the $\beta$ detector and/or a time range since proton impact from 30 to 530 ms. Figure \ref{fig:299_coin} shows the $\gamma$ rays in coincidence with the 6$^{+}$ $\rightarrow$ 4$^{+}$ 299.3~keV transition in $^{132}$Sn. The spectrum illustrates the amount of statistics and the quality of the $\gamma$-ray coincidence spectra.

\begin{figure*}[p]
\includegraphics[width=1.19\textwidth,angle=-90]{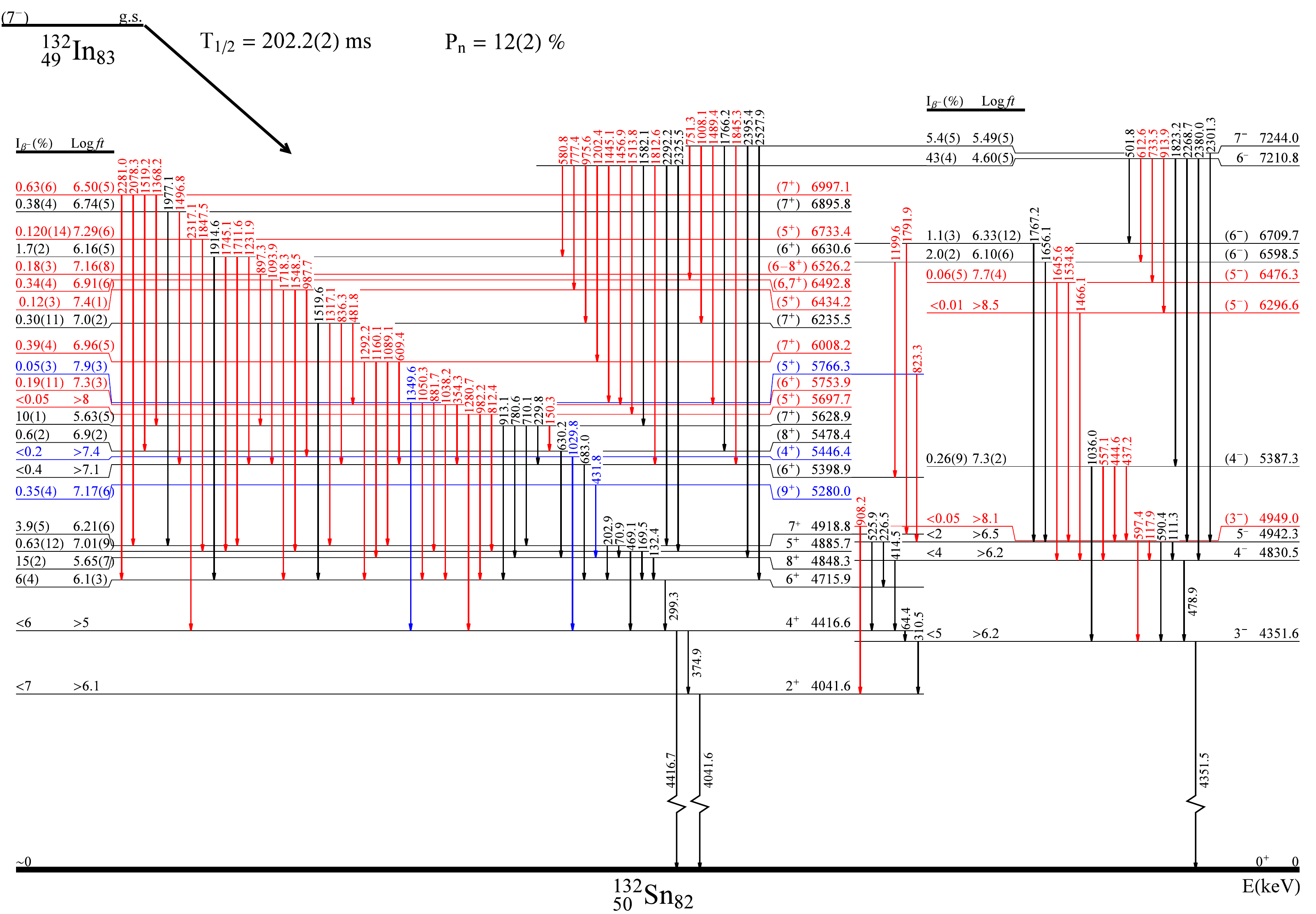}
\caption[width=\textwidth]{\label{fig:LvSc132In} Level scheme of $^{132}$Sn observed following the $\beta$ decay of $^{132}$In. Note that the energy gap from the g.s. to the 2$^+_1$ state is not to scale. The positive-parity states are shown on the left-hand side and the negative parity ones on the right-hand side. The high-lying 7$^-$ and 6$^-$ states feed states of both parities. Levels and transitions previously observed in this decay are colored in black, while those observed for the first time are highlighted in red. Levels previously-known from $^{248}$Cm fission or from $^{133}$In $\beta$-n decay and observed here in the $\beta$-decay of $^{132}$In are colored in blue.}
\end{figure*}

The level scheme of $^{132}$Sn has been greatly expanded with the addition of 57 new $\gamma$ transitions and 11 new levels, observed following the direct $^{132}$In $\beta$ decay. The level scheme is shown in Figure \ref{fig:LvSc132In}. A list of the $\gamma$ rays is provided in Table~\ref{tab:132sngammas}.

The states located at 5766 and 5446 keV were previously observed in the $\beta$-delayed neutron emission of $^{133}$In \cite{Piersa2019}. The assignment is confirmed in this work with the uncovering of new cascades in the $^{132}$In decay involving both levels.

A new level is found at 5754 keV defined by the 1457-1038 keV $\gamma$ cascade. This sequence had already been observed by Fogelberg \textit{et al.} \cite{Fogelberg1994}, however the order of the $\gamma$ rays they proposed was inverted, giving rise to an excited level at 6173 keV with suggested (6$^{+}$) spin-parity. This level has been ruled out in our analysis due to the finding of another $\gamma$-ray that de-excites the 5754-keV state, confirmed by $\gamma$-$\gamma$ coincidences. 

The 5280-keV level was already reported in $^{248}$Cm fission studies \cite{Bhattacharyya2001}, and tentatively identified as the (9$^{+}$) state that arises from the particle-hole $\nu f_{7/2}h^{-1}_{11/2}$ configuration. That level is confirmed in this work by the observation of the de-exciting 431.8-keV transition in delayed coincidence with the 132-keV $\gamma$-ray that de-excites the 8$^{+}$ 4848-keV level. 

The direct $\beta$ feeding to states in $^{132}$Sn was determined from the balance between feeding and de-exciting $\gamma$ rays to each level. The intensities of the $\gamma$ rays were obtained from the measured HPGe singles with neither coincidence nor condition on the time from the impact of protons on target. Theoretical internal conversion coefficients have been  taken from \cite{KIBEDI2008202} if required. The $\beta$ feeding to the states should be understood as upper limits and the log$ft$ values as lower limits due to possible missing transitions.

\subsubsection{$\beta$-delayed neutron branches in $^{132}$In}

The $\beta$-delayed one neutron emission of $^{132}$In has been confirmed by the observation of $\gamma$ rays belonging to the excited structure of $^{131}$Sn. Two transitions have been identified in this decay, specifically, the 2435.0(3) keV and 4273.6(5) keV  with an  absolute intensity  of 0.11(1)$\%$ and 0.04(1)$\%$ respectively.
The P$_n$ value has been obtained from the analysis of the $\gamma$-ray intensities of the $\beta$ and $\beta$-n branches, following the $\beta$ decay of the $^{132}$Sn and $^{131}$Sn daughters. For the $^{132}$Sn $\rightarrow$ $^{132}$Sb decay branch, the intensities of the 5 most intense $\gamma$ rays were considered. The absolute intensities for those transitions were adopted from \cite{PhysRevC.39.1963}. Previous measurements of the $^{131}$Sn $\rightarrow$ $^{131}$Sb decay \cite{Huck1981} were not able to disentangle the decay of the $^{131}$In isomers. Besides, in our analysis we observed that the intensity of the $\gamma$ rays emitted in this decay mainly originate from the population of levels with high spin (11/2, 13/2, 15/2). This indicates that the $^{132}$In isotopes that decay by $\beta$-n mostly feed the 11/2$^{-}$ isomer in $^{131}$Sn directly, which is strongly favored against the 3/2$^{+}$ g.s. due to the angular momentum difference.\par

The total intensity from $^{131}$Sn $\rightarrow$ $^{131}$Sb decay was calculated from the analysis of the $\gamma$ rays from $^{131}$Sb observed in this data set. An absolute intensity of 69(7)$\%$ has been estimated for the most intense transition, of 1226 keV. This value was obtained by making two assumptions: firstly that $^{132}$In $\beta$-n decay mainly populates the (11/2$^{-}$) isomer, which is consistent with our observations, and secondly that no direct intensity was lost due to direct population of $^{131}$Sn to the $^{131}$Sb g.s. This is to be expected in order to be consistent with the first assumption due to the large spin difference (11/2$^{-}$ $\rightarrow$ 7/2$^{+}$ transition). The 7$\%$ uncertainty in the value takes into account these assumptions. \par

Finally, the number of decays calculated for each tin isotope was corrected to account for the movement of the tape at the end of each super-cycle, considering the different half-lives of each isotope and for dead time effects. From this analysis a P$_n$=12(2)$\%$ value was found for the $^{132}$In $\beta$-n decay, which is notably higher than the 6.8(14)$\%$ in \cite{NSR1980LU04} but in good agreement with the 10.7(33)$\%$ from \cite{Rudstam1993}.\par

\LTcapwidth=\columnwidth
\begingroup
\begin{longtable}[]{cccccc}
 
\caption{\label{tab:132sngammas} List of $\gamma$ rays observed in the $\beta$ decay of $^{132}$In to $^{132}$Sn, including transition energies and intensities. The initial and final levels for each connecting transition are also given. }
 \\\hline
 
  E$_{i}$(keV)& J$^{\Pi}_i$& E$_{f}$(keV)& J$^{\Pi}_f$ & E$_{\gamma}$ (keV) & I$_{\gamma}^a$\\\hline\hline
  \endfirsthead
  \caption{(Continued)}\\\hline
    E$_{i}$(keV)& J$^{\Pi}_i$& E$_{f}$(keV)& J$^{\Pi}_f$ & E$_{\gamma}$ (keV) & I$_{\gamma}^a$ \\\hline
  \endhead
  \hline \hline
\multicolumn{6}{l}{$^a$ Relative $\gamma$ intensities normalized to 100 units for the}\\
\multicolumn{6}{l}{\;\,  4$^+ \xrightarrow{}$2$^+$ 375-keV transition. For intensity per 100 }\\
 \multicolumn{6}{l}{\;\, decays of the parent, multiply by 0.56(4).}\\

 \multicolumn{6}{l}{$^b$ Intensity from $\gamma$-$\gamma$ coincidences.}\\ 
 
  
  \endlastfoot
  \hline \multicolumn{6}{c}{\textit{Continued on next table}} \\
\endfoot
 4041.6(3)  & 2$^+$   & 0         & 0$^+$   & 4041.6(3) & 100(11)      \\ \hline
4351.6(3)  & 3$^-$   & 0         & 0$^+$   & 4351.5(3) & 43(5)         \\*  \nopagebreak
           &         & 4041.6(3) & 2$^+$   & 310.5(3)  & 4.2(3)       \\ \hline
4416.6(3)  & 4$^+$   & 0         & 0$^+$   & 4416.7(3) & 16(2)         \\*  \nopagebreak
           &         & 4041.6(3) & 2$^+$   & 374.9(3)  & 100           \\*  \nopagebreak
           &         & 4351.6(3) & 3$^-$   & 64.4(3)   & 1.29(9)      \\ \hline
4715.9(4)  & 6$^+$   & 4416.6(3) & 4$^+$   & 299.3(3)  & 82(6)        \\ \hline
4830.5(4)  & 4$^-$   & 4351.6(3) & 3$^-$   & 478.9(3)  & 45(3)         \\*  \nopagebreak
           &         & 4416.6(3) & 4$^+$   & 414.5(3)  & 0.79(6)      \\ \hline
4848.3(5)  & 8$^+$   & 4715.9(4) & 6$^+$   & 132.4(3)  & 26(2)        \\ \hline
4885.7(5)  & 5$^+$   & 4416.6(3) & 4$^+$   & 469.1(5)  & 2.5(2)        \\*  \nopagebreak
           &         & 4715.9(4) & 6$^+$   & 169.5(4)  & 0.12(5)$^b$  \\ \hline
4918.8(5)  & 7$^+$   & 4715.9(4) & 6$^+$   & 202.9(3)  & 8.0(6)        \\*  \nopagebreak
           &         & 4848.3(5) & 8$^+$   & 70.9(4)   & 1.2(2)$^b$   \\ \hline
4942.4(4)  & 5$^-$   & 4351.6(3) & 3$^-$   & 590.4(3)  & 1.07(8)       \\*  \nopagebreak
           &         & 4416.6(3) & 4$^+$   & 525.9(3)  & 33(2)         \\*  \nopagebreak
           &         & 4715.9(4) & 6$^+$   & 226.5(3)  & 0.67(5)       \\*  \nopagebreak
           &         & 4830.5(4) & 4$^-$   & 111.3(3)  & 2.8(2)       \\ \hline
4949.0(5)  & (3$^-$) & 4830.5(4) & 4$^-$   & 117.9(5)  & 0.012(4)$^b$  \\*  \nopagebreak
           &         & 4351.6(3) & 3$^-$   & 597.4(6)  & 0.11(3)$^b$   \\*  \nopagebreak
           &         & 4041.6(3) & 2$^+$   & 908.2(4)  & 0.016(6)$^b$ \\ \hline
5280.0(6)) & (9$^+$) & 4848.3(5) & 8$^+$   & 431.8(4)  & 0.63(6)$^b$  \\ \hline
5387.3(3)  & (4$^-$) & 4351.6(3) & 3$^-$   & 1036.0(3) & 1.29(11)      \\*  \nopagebreak
           &         & 4830.5(4) & 4$^-$   & 557.1(4)  & 0.13(2)       \\*  \nopagebreak
           &         & 4942.4(4) & 5$^-$   & 444.6(4)  & 0.23(3)       \\*  \nopagebreak
           &         & 4949.0(5) & (3$^-$) & 437.2(4)  & 0.23(2)      \\ \hline
5398.9(5)  & (6$^+$) & 4715.9(4) & 6$^+$   & 683.0(3)  & 3.9(3)       \\ \hline
5446.4(5)  & (4$^+$) & 4416.6(3) & 4$^+$   & 1029.8(4) & 0.26(3)      \\ \hline
5478.4(6)  & (8$^+$) & 4848.3(5) & 8$^+$   & 630.2(3)  & 4.6(3)       \\ \hline
5628.9(3)  & (7$^+$) & 4715.9(4) & 6$^+$   & 913.1(3)  & 12.2(10)      \\*  \nopagebreak
           &         & 4848.3(5) & 8$^+$   & 780.6(3)  & 5.0(4)        \\*  \nopagebreak
           &         & 4918.8(5) & 7$^+$   & 710.1(3)  & 2.0(2)        \\*  \nopagebreak
           &         & 5398.9(5) & (6$^+$) & 229.8(3)  & 0.65(5)       \\*  \nopagebreak
           &         & 5478.4(6) & (8$^+$) & 150.3(3)  & 0.29(3)$^b$  \\ \hline
5697.7(3)  & (5$^+$) & 4416.6(3) & 4$^+$   & 1280.7(3) & 0.134(13)     \\*  \nopagebreak
           &         & 4715.9(4) & 6$^+$   & 982.2(5)  & 0.08(2)$^b$   \\*  \nopagebreak
           &         & 4885.7(5) & 5$^+$   & 812.4(6)  & 0.05(2)$^b$  \\ \hline
5753.9(4)  & (6$^+$) & 4715.9(4) & 6$^+$   & 1038.2(3) & 1.8(2)        \\*  \nopagebreak
           &         & 5398.9(5) & (6$^+$) & 354.3(3)  & 0.58(4)      \\ \hline
5766.3(3)  & (5$^+$) & 4416.6(3) & 4$^+$   & 1349.6(3) & 0.27(2)       \\*  \nopagebreak
           &         & 4715.9(4) & 6$^+$   & 1050.3(4) & 0.122(10)     \\*  \nopagebreak
           &         & 4885.7(5) & 5$^+$   & 881.7(3)  & 0.114(14)     \\*  \nopagebreak
           &         & 4942.4(4) & 5$^-$   & 823.2(7)  & 0.068(5)     \\ \hline
6008.2(4)  & (7$^+$) & 4715.9(4) & 6$^+$   & 1292.2(5) & 0.042(7)$^b$  \\*  \nopagebreak
           &         & 4848.3(5) & 8$^+$   & 1160.1(3) & 0.163(13)     \\*  \nopagebreak
           &         & 4918.8(5) & 7$^+$   & 1089.1(4) & 0.15(2)       \\*  \nopagebreak
           &         & 5398.9(5) & (6$^+$) & 609.4(4)  & 0.37(5)      \\ \hline
6235.5(3)  & (7$^+$) & 4715.9(4) & 6$^+$   & 1519.6(3) & 0.54(5)$^b$   \\*  \nopagebreak
           &         & 4918.8(5) & 7$^+$   & 1317.1(3) & 0.31(3)       \\*  \nopagebreak
           &         & 5398.9(5) & (6$^+$) & 836.3(4)  & 0.64(6)       \\*  \nopagebreak
           &         & 5753.9(4) & (6$^+$) & 481.8(3)  & 0.53(6)      \\ \hline
 6296.6(5) & (5$^-$)   & 4830.5(4) & 4$^-$     & 1466.1(3) & 0.120(15)$^b$ \\ \hline
6434.2(3) & (5$^+$)   & 4715.9(4) & 6$^+$     & 1718.3(4) & 0.091(10)      \\*  \nopagebreak
          &           & 4885.7(5) & 5$^+$     & 1548.5(4) & 0.14(2)        \\*  \nopagebreak
          &           & 5446.4(5) & (4$^+$)   & 987.7(3)  & 0.14(4)$^b$   \\ \hline
6476.3(4) & (5$^-$)   & 4830.5(4) & 4$^-$     & 1645.6(3) & 0.40(7)        \\*  \nopagebreak
          &           & 4942.4(4) & 5$^-$     & 1534.8(5) & 0.085(9)      \\ \hline
6492.8(3) & (6,7$^+$) & 5398.9(5) & (6$^+$)   & 1093.9(3) & 0.73(5)       \\ \hline
6526.2(5) & (6-8$^+$) & 5628.9(3) & (7$^+$)   & 897.3(3)  & 0.32(4)       \\ \hline
6598.5(5) & (6$^-$)   & 4942.4(4) & 5$^-$     & 1656.1(3) & 4.0(3)         \\*  \nopagebreak
          &           & 5398.9(5) & (6$^+$)   & 1199.6(5) & 0.036(10)$^b$ \\ \hline
6630.6(4) & (6$^+$)   & 4715.9(4) & 6$^+$     & 1914.6(3) & 2.5(2)         \\*  \nopagebreak
          &           & 4885.7(5) & 5$^+$     & 1745.1(5) & 0.23(5)        \\*  \nopagebreak
          &           & 4918.8(5) & 7$^+$     & 1711.6(3) & 0.24(2)        \\*  \nopagebreak
          &           & 5398.9(5) & (6$^+$)   & 1231.9(4) & 0.27(3)$^b$   \\ \hline
6709.7(4) & (6$^-$)   & 4918.8(5) & 7$^+$     & 1791.0(3) & 0.42(4)        \\*  \nopagebreak
          &           & 4942.4(4) & 5$^-$     & 1767.2(3) & 3.2(3)$^b$    \\ \hline
6733.4(4) & (5$^+$)   & 4416.6(3) & 4$^+$     & 2317.1(4) & 0.098(10)      \\*  \nopagebreak
          &           & 4885.7(5) & 5$^+$     & 1847.5(3) & 0.12(2)$^b$   \\ \hline
6895.8(4) & (7$^+$)   & 4918.8(5) & 7$^+$     & 1977.1(3) & 0.31(3)        \\*  \nopagebreak
          &           & 5398.9(5) & (6$^+$)   & 1496.8(3) & 0.37(3)       \\ \hline
6997.1(3) & (7$^+$)   & 4715.9(4) & 6$^+$     & 2281.0(3) & 0.35(3)        \\*  \nopagebreak
          &           & 4918.8(5) & 7$^+$     & 2078.3(3) & 0.22(2)        \\*  \nopagebreak
          &           & 5478.4(6) & (8$^+$)   & 1519.2(5) & 0.14(2)$^b$    \\*  \nopagebreak
          &           & 5628.9(3) & (7$^+$)   & 1368.2(5) & 0.43(5)$^b$   \\ \hline
7210.8(3) & 6$^-$     & 4830.5(4) & 4$^-$     & 2380.0(3) & 38(3)          \\*  \nopagebreak
          &           & 4885.7(5) & 5$^+$     & 2325.5(3) & 0.85(6)        \\*  \nopagebreak
          &           & 4918.8(5) & 7$^+$     & 2292.2(3) & 1.04(10)       \\*  \nopagebreak
          &           & 4942.4(4) & 5$^-$     & 2268.7(3) & 27(2)          \\*  \nopagebreak
          &           & 5387.3(3) & (4$^-$)   & 1823.2(3) & 1.41(10)       \\*  \nopagebreak
          &           & 5398.9(5) & (6$^+$)   & 1812.6(6) & 0.12(1)       \\* \nopagebreak
          &           & 5628.9(3) & (7$^+$)   & 1582.1(3) & 1.16(9)       \\* \nopagebreak
          &           & 5697.7(3) & (5$^+$)   & 1513.8(3) & 0.34(3)       \\* \nopagebreak
          &           & 5753.9(4) & (6$^+$)   & 1456.9(3) & 1.44(11)      \\* \nopagebreak
          &           & 5766.3(3) & (5$^+$)   & 1445.1(3) & 0.49(4)       \\* \nopagebreak
          &           & 6008.2(4) & (7$^+$)   & 1202.4(6) & 0.024(7)$^b$  \\* \nopagebreak
          &           & 6235.5(3) & (7$^+$)   & 975.6(3)  & 1.4(2)$^b$    \\* \nopagebreak
          &           & 6296.6(5) & (5$^-$)   & 913.9(4)  & 0.121(14)$^b$ \\* \nopagebreak
          &           & 6434.2(3) & (5$^+$)   & 777.4(3)  & 0.15(2)$^b$   \\* \nopagebreak
          &           & 6476.3(4) & (5$^-$)   & 733.5(3)  & 0.38(3)       \\* \nopagebreak
          &           & 6598.5(5) & (6$^-$)   & 612.6(3)  & 0.42(3)       \\* \nopagebreak
          &           & 6709.7(4) & (6$^-$)   & 501.8(4)  & 1.7(4)        \\* \nopagebreak
          &           & 6630.6(4) & (6$^+$)   & 580.8(5)  & 0.09(2)$^b$   \\ \hline
7244.0(3) & 7$^-$     & 4715.9(4) & 6$^+$     & 2527.9(3) & 1.47(10)      \\* \nopagebreak
          &           & 4848.3(5) & 8$^+$     & 2395.4(3) & 2.3(2)        \\* \nopagebreak
          &           & 4942.4(4) & 5$^-$     & 2301.3(3) & 2.0(2)        \\* \nopagebreak
          &           & 5398.9(5) & (6$^+$)   & 1845.3(3) & 0.46(3)       \\* \nopagebreak
          &           & 5478.4(6) & (8$^+$)   & 1766.2(3) & 3.2(3)$^b$    \\* \nopagebreak
          &           & 5753.9(4) & (6$^+$)   & 1489.4(4) & 0.084(11)     \\* \nopagebreak
          &           & 6235.5(3) & (7$^+$)   & 1008.1(5) & 0.11(2)       \\* \nopagebreak
          &           & 6492.8(3) & (6,7$^+$) & 751.3(3)  & 0.124(12)     \\ 

 \end{longtable}
 \endgroup

Given the two-neutron separation energy in the $^{132}$Sn S$_{2n}$($^{132}$Sn)=12557.0(27) keV \cite{Wang2017}, there is a 1583 keV energy window, within Q$_{\beta}$($^{132}$In), that makes the decay via a $\beta$-delayed two neutron branch possible. We have searched for $\gamma$ rays belonging to the \textit{A} = 130 mass chain. Nevertheless, no evidence has been found that would point to the existence of a $\beta$-2n decay branch in $^{132}$In. This is consistent with expectations, since the only levels in $^{130}$Sn that could be populated are the 0$^{+}$ g.s. and 2$^{+}$ at 1221 keV \cite{NDS130}. Assuming a (7$^{-}$) spin-parity assignment for the $^{132}$In g.s. \cite{Jungclaus2016}, this decay will be highly suppressed.

\subsection{$\beta$-delayed neutron decay of $^{133}$In}

The large Q$_{\beta}$=13.4(2) MeV value along with the low neutron separation energy in $^{133}$Sn, S$_n$=2.399(3) MeV \cite{Wang2017}, favor the $^{133}$In decay via $\beta$-delayed neutron emission to $^{132}$Sn. This gives rise to the large P$_n$ values for both $^{133}$In $\beta$-decaying isomers \cite{Piersa2019}.
The lower spin of the $^{133}$In (9/2$^{+}$) ground state and (1/2$^{-}$) isomer, in comparison with $^{132}$In (7$^-$), is expected to favor the population of low spin p-h excited states that are not fed in the $\beta$ decay of $^{132}$In due to the large spin of the parent (7$^{-}$). The population of $^{132}$Sn excited levels in the $\beta$ decay of $^{133}$In was already reported in \cite{Piersa2019}, where the $\beta$ decay of $^{133}$In was investigated focusing on the excited structure of $^{133}$Sn. In the present work, we concentrate on results of the $\beta$-n decay of $^{133}$In to $^{132}$Sn, discuss excited states in $^{132}$Sn and report new transitions following the $\beta$-n decay branch of $^{133}$In.\par

\subsubsection{Feeding of excited states in $^{132}$Sn}

Gamma rays emitted after the $\beta$ decay of $^{133}$In can be clearly distinguished from the background by their time distribution following the impact of protons on target. Nevertheless, the decay curve does not allow the separation of the transitions that belong to $^{132}$Sn from those of $^{133}$Sn, neither from the background induced by $\beta$-delayed neutrons from $^{133}$In $\beta$-n decay.

\begin{figure}[H]
\includegraphics[width=\columnwidth]{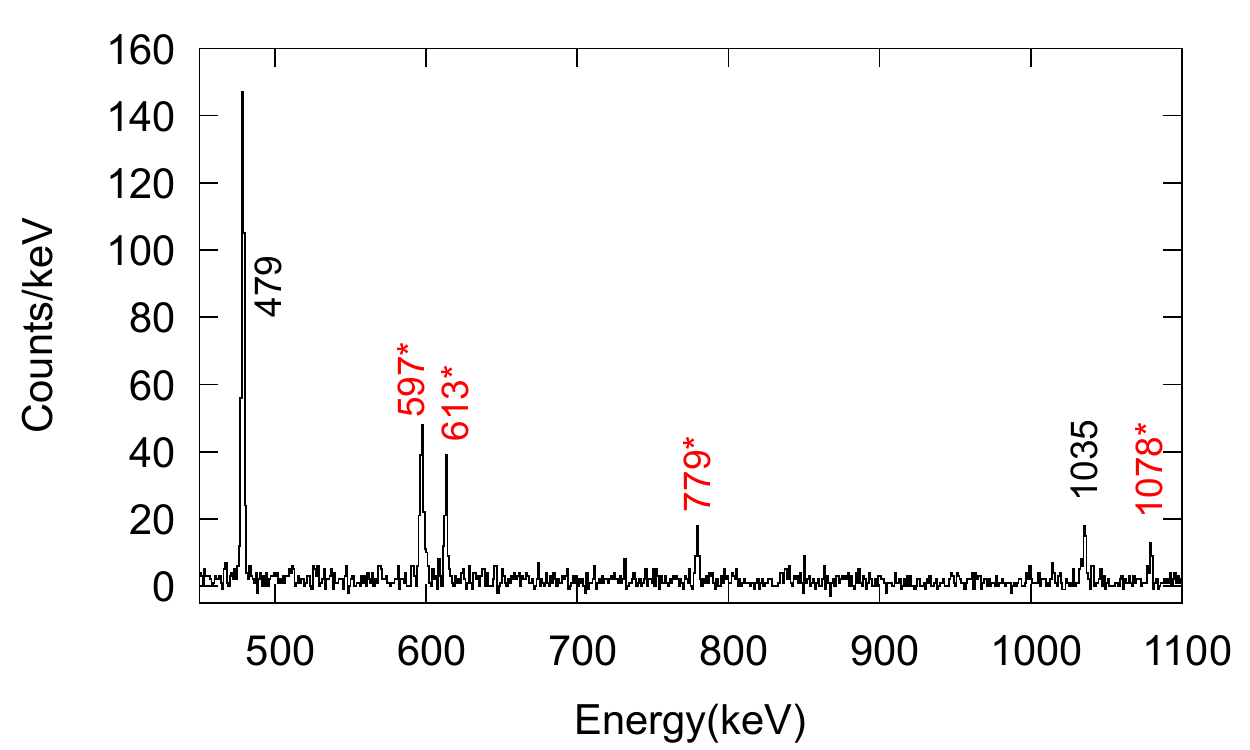}
\caption{\label{fig:4351_gate_gg} $\gamma$-$\gamma$ spectrum observed in the $\beta$ decay of $^{133}$In, gated on the 4352-keV transition in $^{132}$Sn. The contribution from Compton events beneath the 4352-keV peak 
has been subtracted. Newly-observed $\gamma$ transitions are labeled in red with an asterisk.}
\end{figure}

The identification of new $\gamma$ rays that belong to $^{132}$Sn is based on $\gamma$-$\gamma$ coincidences. Gamma rays with energies below 4 MeV are always a part of a cascade since they cannot directly feed the g.s. 
The analysis allows to identify several levels and $\gamma$ rays in $^{132}$Sn that are not observed in the $^{132}$In decay, see Figure \ref{fig:4351_gate_gg}. Among them we confirm the states  at 4965, 5131, 5431 and 5790 keV.

\LTcapwidth=\columnwidth
\begin{longtable}[]{p{0.143\columnwidth}p{0.06\columnwidth}p{0.143\columnwidth}p{0.06\columnwidth}p{0.148\columnwidth}p{0.150\columnwidth}p{0.150\columnwidth}}
\caption{\label{tab:133In_gamma_intensities}List of observed states in $^{132}$Sn following the $\beta$-n decay of the $^{133}$In (9/2$^+$) g.s. (>95\% pure), labeled $^{133g}$In, and the beam with enhanced content of the (1/2$^-$) isomeric state, with a contamination of $\sim$30\% of $^{133g}$In, labeled $^{133m}$In.}\\ \hline
\hfil E$_{i}$   &\hfil J$^{\Pi}_i$ &\hfil E$_{f}$ &\hfil J$^{\Pi}_f$ &\hfil E$_{\gamma}$ &\hfil I$_{rel}^c$   &\hfil I$_{rel}^d$          \\
\hfil  (keV) &             &  (keV)   &             & \hfil   (keV)    & \hfil  $^{133g}$In& \hfil       $^{133m}$In             \\ \hline \hline
\endfirsthead
\caption{(Continued)}\\ \hline
\hfil E$_{i}$     &\hfil J$^{\Pi}_i$ &\hfil E$_{f}$    &\hfil J$^{\Pi}_f$ &\hfil E$_{\gamma}$  &\hfil I$_{rel}^c$  &\hfil I$_{rel}^d$                   \\
  (keV) &             &\hfil  (keV)   &             &\hfil    (keV)    &  \hfil $^{133g}$In&  \hfil      $^{133m}$In             \\ \hline\hline
\endhead

\multicolumn{7}{l}{$^a$ Intensity obtained from $\gamma$-$\gamma$ coincidences.}\\
\multicolumn{7}{l}{$^b$ Not observed in this decay, intensity calculated from}\\
\multicolumn{7}{l}{\;\, $^{132}$In decay data.}\\

\multicolumn{7}{l}{$^c$ Relative $\gamma$ intensities normalized to 100 units for the}\\
\multicolumn{7}{l}{\;\,4042-keV transition.}\\
\multicolumn{7}{l}{\;\,For intensity per 100 decays multiply by 0.049(5).}\\

\multicolumn{7}{l}{$^d$ Relative $\gamma$ intensities normalized to 100 units for the}\\
\multicolumn{7}{l}{\;\,4042-keV transition.  }\\
\multicolumn{7}{l}{\;\,For intensity per 100 decays multiply by 0.043(5).}\\

\endlastfoot

  \hline \multicolumn{7}{c}{\textit{Continued on next table}} \\
\endfoot
\hline
4041.6(3) &\hfil 2$^+$       & 0         &\hfil 0$^+$       & 4041.6(3)    &\hfil 100          &\hfil 100                              \\ \hline
4351.6(3) &\hfil 3$^-$       & 0         &\hfil 0$^+$       & 4351.5(3)    &\hfil 54(8)        & \hfil 70(11)                           \\*  \nopagebreak
          &             & 4041.6(3) &\hfil 2$^+$       & 310.5(3)     &\hfil 7.0(9)       &\hfil 12(2)                            \\ \hline
4416.6(3) &\hfil 4$^+$       & 0         &\hfil 0$^+$       & 4416.7(3)    &\hfil 13(2)        &\hfil 9(2)                             \\*  \nopagebreak
          &             & 4041.6(3) &\hfil 2$^+$       & 374.9(3)     &\hfil 74(9)        &\hfil 39(6)                            \\*  \nopagebreak
          &             & 4351.6(3) &\hfil 3$^-$       & 64.4(3)      & 0.96(11)$^b$ &\hfil 0.51(7)$^b$                      \\ \hline
4715.9(4) &\hfil 6$^+$       & 4416.6(3) &\hfil 4$^+$       & 299.3(3)     &\hfil 31(4)        &\hfil 13(2)                            \\ \hline
4830.5(4) &\hfil 4$^-$       & 4351.6(3) &\hfil 3$^-$       & 478.9(3)     &\hfil 18(2)        &\hfil 18(2)                            \\*  \nopagebreak
          &             & 4416.6(3) &\hfil 4$^+$       & 414.5(3)     &\hfil 0.9(5)       &\hfil 0.21(8)$^a$                      \\ \hline
4848.3(5) &\hfil 8$^+$       & 4715.9(4) &\hfil 6$^+$       & 132.4(3)     &\hfil 4.0(6)       &\hfil \textless2$^a$   \\ \hline
4885.7(5) &\hfil 5$^+$       & 4416.6(3) &\hfil 4$^+$       & 469.1(5)     &\hfil11.1(14)     &\hfil 7(2)                             \\*  \nopagebreak
          &             & 4715.9(4) &\hfil 6$^+$       & 169.5(4)     &\hfil 1.5(7)$^a$   &\hfil \textless1.6$^a$ \\ \hline
4918.8(5) &\hfil 7$^+$       & 4715.9(4) &\hfil 6$^+$       & 202.9(3)     &\hfil 5.4(8)       &\hfil 3(2)                             \\*  \nopagebreak
          &             & 4848.3(5) &\hfil 8$^+$       & 70.9(4)      &\hfil 0.8(1)$^b$   &\hfil 0.4(2)$^b$                       \\ \hline
4942.4(4) &\hfil 5$^-$       & 4351.6(3) &\hfil 3$^-$       & 590.4(3)     &\hfil 0.7(6)$^a$   &\hfil 0.22(3)$^b$                      \\*  \nopagebreak
          &             & 4416.6(3) &\hfil 4$^+$       & 525.9(3)     &\hfil 10.1(14)     &\hfil 6.6(10)                          \\*  \nopagebreak
          &             & 4715.9(4) &\hfil 6$^+$       & 226.5(3)     & 0.21(3)$^b$  &\hfil 0.14(2)$^b$                      \\*  \nopagebreak
          &             & 4830.5(4) &\hfil 4$^-$       & 111.3(3)     &\hfil 1.0(3)       &\hfil 0.56(9)$^b$                      \\ \hline
4949.0(5) & (3$^-$)     & 4830.5(4) &\hfil 4$^-$       & 117.9(5)     &\hfil 1.4(4)       &\hfil 1.0(7)                           \\*  \nopagebreak
          &             & 4351.6(3) &\hfil 3$^-$       & 597.4(6)     &\hfil 8(2)$^a$     &\hfil 12(4)$^a$                        \\*  \nopagebreak
          &             & 4041.6(3) &\hfil 2$^+$       & 908.2(4)     &\hfil 1.5(3)       &\hfil \textless 8       \\ \hline
4965.3(7) & (3$^+$)     & 4416.6(3) &\hfil 4$^+$       & 549.0(4)     &\hfil 1.0(6)       &\hfil 1.9(8)$^a$                       \\*  \nopagebreak
          &             & 4351.6(3) &\hfil 3$^-$       & 613.5(5)     &\hfil 3.8(5)       &\hfil 13(5)$^a$                        \\*  \nopagebreak
          &             & 4041.6(3) &\hfil 2$^+$       & 923.8(7)     &\hfil 2.4(5)       &\hfil 3.2(12)                          \\ \hline
5131.2(6) & (2$^-$)     & 4351.6(3) &\hfil 3$^-$       & 779.2(4)     &\hfil 2.3(4)       &\hfil 14(3)                            \\*  \nopagebreak
          &             & 0         &\hfil 0$^+$       & 5131.9(8)    &\hfil 1.1(5)       &\hfil 4(3)                             \\ \hline
5387.3(3) & (4$^-$)     & 4351.6(3) &\hfil 3$^-$       & 1036.0(3)    &\hfil 3.9(5)       &\hfil 3.9(12)                          \\*  \nopagebreak
          &             & 4830.5(4) &\hfil 4$^-$       & 557.1(4)     &\hfil 0.39(6)$^b$  &0.39(13)$^b$                     \\*  \nopagebreak
          &             & 4942.4(4) &\hfil 5$^-$       & 444.6(4)     & 0.71(14)$^b$ &\hfil 0.7(2)$^b$                       \\*  \nopagebreak
          &             & 4949.0(5) & (3$^-$)     & 437.2(4)     &\hfil 0.7(2)$^b$   &\hfil 0.7(2)$^b$                       \\ \hline
 5398.9(5) & (6$^+$)     & 4715.9(4) &\hfil 6$^+$       & 683.0(3)     &\hfil 4.7(7)       &\hfil 2.2(19)                           \\ \hline
5431.4(7) & (3)         & 4041.6(3) &\hfil 2$^+$       & 1390.3(14)   &\hfil 2.0(7)       &\hfil 4.8(8)                            \\*  \nopagebreak
          &             & 4351.6(3) &\hfil 3$^-$       & 1078.9(7)    &\hfil 2.4(8)$^a$   &\hfil 4(2)                              \\*  \nopagebreak
          &             & 4416.6(3) &\hfil 4$^+$       & 1015.4(10)   &\hfil 2.8(7)$^a$   &\hfil 1.1(8)$^a$                        \\ \hline
5446.4(5) & (4$^+$)     & 4416.6(3) &\hfil 4$^+$       & 1029.8(4)    &\hfil 4.6(6)       &\hfil 2.9(12)                           \\ \hline
5478.4(6) & (8$^+$)     & 4848.3(5) &\hfil 8$^+$       & 630.2(3)     &\hfil 3.2(7)       &\hfil 3.2(13)                           \\ \hline
5628.9(3) & (7$^+$)     & 4715.9(4) &\hfil 6$^+$       & 913.1(3)     &\hfil 1.9(3)       &\hfil \textless 1$^a$    \\*  \nopagebreak
          &             & 4848.3(5) &\hfil 8$^+$       & 780.6(3)     & 0.79(13)$^b$ &\hfil \textless 0.3$^b$  \\*  \nopagebreak
          &             & 4918.8(5) &\hfil 7$^+$       & 710.1(3)     &\hfil 0.31(5)$^b$  &\hfil \textless 0.2$^b$  \\*  \nopagebreak
          &             & 5398.9(5) & (6$^+$)     & 229.8(3)     &\hfil 0.10(2)$^b$  &\hfil \textless 0.05$^b$ \\*  \nopagebreak
          &             & 5478.4(6) &(8$^+$)     & 150.3(3)     & 0.047(8)$^b$ &\hfil \textless 0.03$^b$ \\ \hline
5697.7(3) & (5$^+$)     & 4416.6(3) &\hfil 4$^+$       & 1280.7(2)    &\hfil 0.8(3)       &\hfil 2.1(17)$^b$                       \\*  \nopagebreak
          &             & 4715.9(4) &\hfil 6$^+$       & 982.2(3)     &\hfil 0.8(4)       &\hfil 1.3(9)$^a$                        \\*  \nopagebreak
          &             & 4885.7(5) &\hfil 5$^+$       & 812.4(6)     &\hfil 0.5(4)$^b$   &\hfil 0.7(7)$^b$                        \\ \hline
5753.9(4) & (6$^+$)     & 4715.9(4) &\hfil 6$^+$       & 1038.2(3)    &\hfil 2.0(6)$^a$   &\hfil 0.8(7)$^a$                        \\*  \nopagebreak
          &             & 5398.9(5) & (6$^+$)     & 354.3(3)     &\hfil 0.6(2)$^b$   &\hfil 0.3(2)$^b$                        \\ \hline
5766.2(3) & (5$^+$)     & 4416.6(3) &\hfil 4$^+$       & 1349.6(3)    &\hfil 1.4(10)      &\hfil 2.3(15)$^a$                       \\*  \nopagebreak
          &             & 4715.9(4) &\hfil 6$^+$       & 1050.3(4)    &\hfil 0.9(4)$^a$   &\hfil 0.46(7)                           \\*  \nopagebreak
          &             & 4885.7(5) &\hfil 5$^+$       & 881.7(3)     &\hfil 0.9(3)       &\hfil 0.24(6)                           \\*  \nopagebreak
          &             & 4942.4(4) &\hfil 5$^-$       & 823.2(7)     &\hfil 0.4(2)$^b$   &\hfil 0.26(4)$^a$                       \\ \hline
5790.4(6) & (4$^+$)     & 4416.6(3) &\hfil 4$^+$       & 1373.9(4)    &\hfil 2.7(3)       &\hfil 3.3(13)$^a$                       \\*  \nopagebreak
          &             & 4351.6(3) &\hfil 3$^-$       & 1438.5(4)    &\hfil 0.8(3)       &\hfil 1.0(6)                            \\ \hline
6296.6(5) &(5$^-$)     & 4830.5(4) &\hfil 4$^-$       & 1466.1(3)    &\hfil 1.9(11)$^a$  &\hfil \textless 0.5$^a$  \\ \hline
6476.3(4) & (5$^-$)     & 4830.5(4) &\hfil 4$^-$       & 1645.6(3)    &\hfil 1.6(10)$^a$  &\hfil \textless 0.5$^a$  \\*  \nopagebreak
          &             & 4942.4(4) &\hfil 5$^-$       & 1534.8(5)    &\hfil 0.3(2)$^b$   &\hfil \textless 0.1$^b$  \\ \hline

\end{longtable}

\begin{figure*}[p]
\centering
\includegraphics[width=1.20\textwidth,angle=-90]{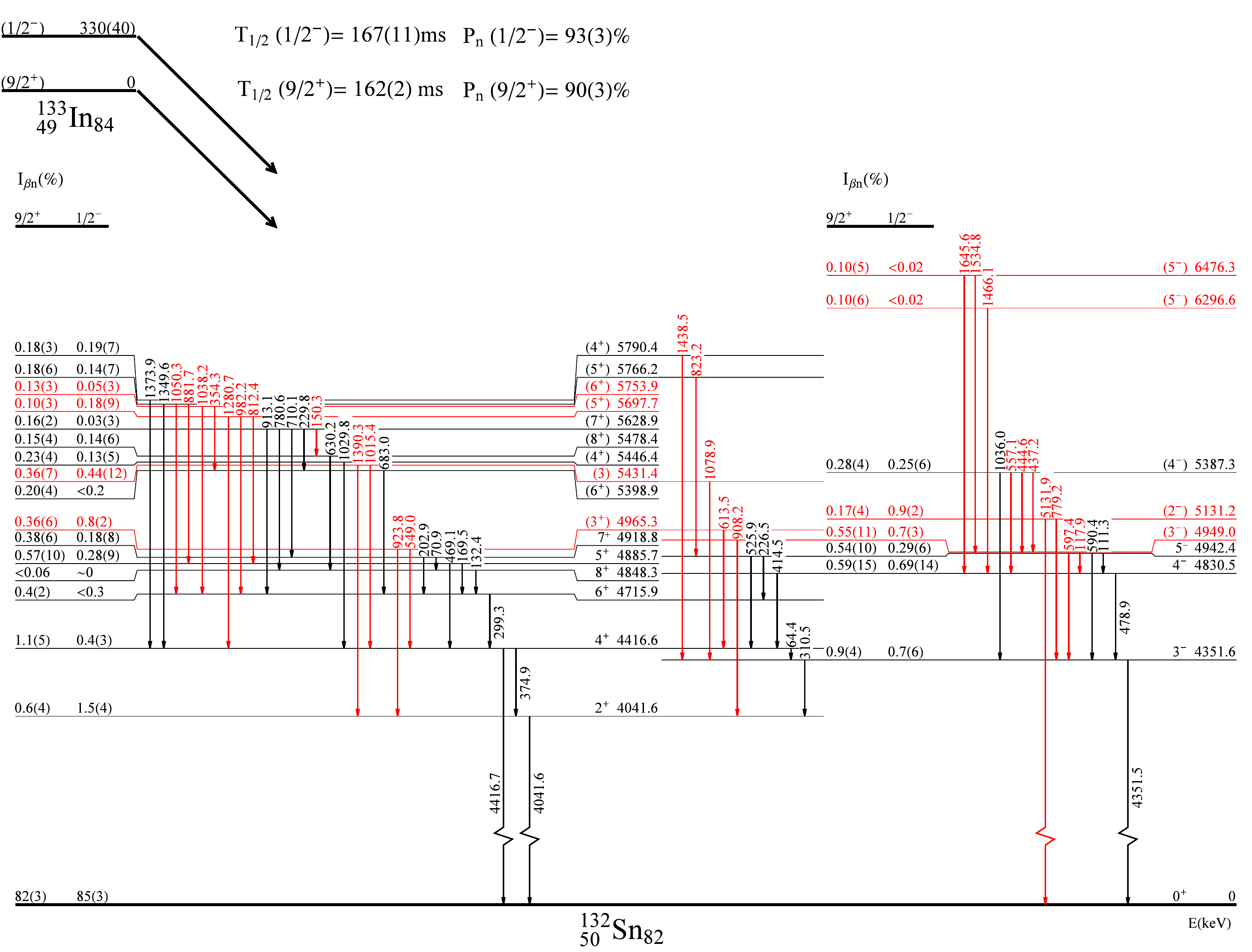}
\caption[width=\textwidth]{\label{fig:LvSc133In} Level scheme of $^{132}$Sn observed following the $\beta$-n decay of the $^{133}$In (9/2$^{+}$) ground state and (1/2$^{-}$) isomer. Note that the energy gap from the g.s. to the 2$^+_1$ state is not to scale. The positive parity states on the left-hand side and the negative parity states to the right.  The isomer observed feeding to levels in $^{132}$Sn is provided separately for each decay. Note contributions coming from beam impurity cannot be excluded in the (1/2$^-$) decay. Levels and transitions observed for the first time in the $\beta$-decay of $^{133}$In are highlighted in red. }
\end{figure*}

The low spin of the states to which they can de-excite suggests a small spin value (2-4) for these levels, which makes them good candidates for the remaining particle-hole states with low spin expected in this energy range. In Figure \ref{fig:LvSc133In} the level-scheme of $^{132}$Sn in the $\beta$-n decay of $^{133}$In is depicted. 
The direct feeding to each level (I$_{\beta-n}$) is measured by analyzing the $\gamma$-ray intensities calculated separately for each isomer. The indium beams for each isomer were separated taking advantage of the isomer selectivity provided by RILIS. However the separation was not complete, and the $^{133m}$In beam contained a contribution of $\sim$30\% of $^{133g}$In \cite{Piersa2019}. The amount of $^{133m}$In in the $^{133g}$In beam is below 5\%. The total feeding has been calculated using the intensities of the $\gamma$-rays emitted by the daughters. The observed states in $^{132}$Sn following  the $\beta$-n decay of the $^{133}$In (9/2$^+$) g.s. and (1/2$^-$) isomeric state are listed in Table \ref{tab:133In_gamma_intensities}.

\subsubsection{High energy $\gamma$-rays}
Another interesting feature observed in the $\beta$-decay of $^{133}$In is the presence of several $\gamma$ rays at very high energies, above 5 MeV, with a time behavior compatible with the decay of $^{133g,m}$In. The existence of those $\gamma$ rays was already reported in \cite{Piersa2019}, where the 6088-keV transition was assigned to $^{133}$Sn. Some other $\gamma$-rays were discussed in \cite{Piersa2019} as being emitted in the decay of $^{133}$In, however the lack of $\gamma$-$\gamma$ coincidences does not allow to identify the daughter tin isotopes they belong to.\par

\begin{figure}[ht!]
\includegraphics[width=\columnwidth]{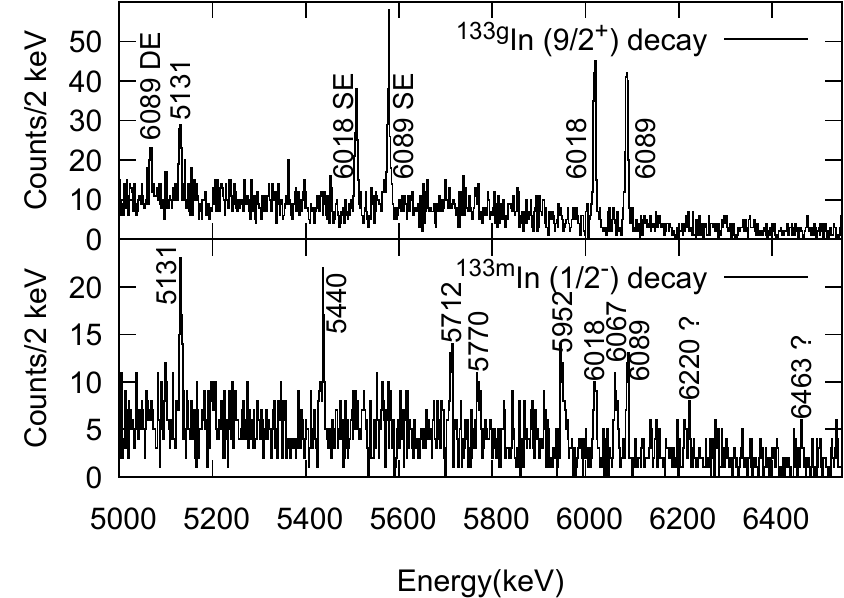}
\caption{\label{fig:high_energy} Beta-gated $\gamma$-ray spectra from the $\beta$ decay of $^{133}$In isomers highlighting the energy range above 5 MeV. Only the events recorded in the time window from 10 to 600 ms since the arrival of the proton pulse are used.}
\end{figure}

The high energy $\gamma$ lines observed in the $^{133g,m}$In decays are shown in Figure \ref{fig:high_energy}. As it can be seen, the observed peaks differ notably depending on the selected $\beta$-decaying indium state. In the decay of the $^{133g}$In (9/2$^+$) g.s. there are two predominant $\gamma$-rays, the one at 6088 keV mentioned above, and another one at 6019 keV. 
Although the 6019-keV peak has the same energy as a transition in $^{57}$Fe produced by the neutron background as discussed in \cite{Piersa2019} its
intensity is only a small fraction of the total $\gamma$-ray intensity from the excited 7647-keV level in $^{57}$Fe, while the other, more intense transitions, are not observed. 
Therefore the 6019-keV transition is likely to be emitted following the $\beta$-decay of $^{133}$In, and predominantly from the (9/2$^+$) g.s.\par

In the decay of the $^{133m}$In (1/2$^{-}$), the 6088- and 6019-keV transitions are suppressed, but several other peaks, which are absent in the decay of the $^{133g}$In, can be identified. Those peaks appear at the energies of 5440, 5712, 5770, 5952, and 6067 keV. Two more tentative peaks,
at the detection limit of the HPGe detectors, are seen at 6220 and 6463 keV. Some of these $\gamma$ lines may be compatible with escape peaks from other $\gamma$ rays (for instance single and double escape peaks from 6463 keV), but it is not possible to make a consistent identification for all of the energies. 
It is interesting to observe transitions having this energy from the decay of the (1/2$^{-}$) isomer, as there are unidentified members of particle-hole multiplets in $^{132}$Sn, such as $\nu$p$_{3/2}$d$^{-1}_{3/2}$, $\pi$g$_{7/2}$g$^{-1}_{9/2}$ and $\nu$p$_{3/2}$s$^{-1}_{1/2}$, that can give rise to low spin levels. It is very unlikely to populate them in the decay of $^{132}$In with (7$^{-}$). However, the feeding of such levels would be strongly favored in the $\beta$-n decay of the (1/2$^{-}$) state in $^{133m}$In. All of this points towards these transitions likely originating from the de-excitation of such p-h multiplet states.

It is worth mentioning that a 5131-keV peak can be seen in both the $^{133g}$In and $^{133m}$In decays. This $\gamma$ ray has been firmly identified to belong to $^{132}$Sn since its energy perfectly matches the de-excitation of the new 5131-keV level proposed in this work. 
In addition, the existence of a transition to the 0$^{+}$ g.s. supports the tentative assignment of this level to the 2$^{-}$ state of the $\nu$f$_{7/2}$d$^{-1}_{3/2}$ multiplet.\par

For the sake of completeness, the $\gamma$-rays observed in the $^{133}$In decay that have not been assigned to any decay branch are listed in Table~\ref{tab:Unassigned_gamma_rays}.

\begin{table}[ht!]
\centering
\caption{Gamma rays observed in the $^{133}$In decay that could not been assigned to any specific decay branch. Intensities are given relative to the 4042-keV $\gamma$-ray intensity. The label $^{133g}$In refers to the $^{133}$In (9/2$^+$) g.s., with an estimated purity above 95\%, while the label $^{133m}$In is used for the beam with enhanced content of the (1/2$^-$) isomeric state, with a contamination of $\sim$30\% of $^{133g}$In.
}
     
\begin{tabular}{p{0.13\textwidth}p{0.13\textwidth}p{0.13\textwidth}}
\hline 
\hfil E$_{\gamma}$  & \hfil I$_{rel}^a$    & \hfil I$_{rel}^b$ \\
\hfil(keV) &\hfil $^{133g}$In (9/2$^{+}$) &\hfil $^{133m}$In (1/2$^{-}$)\\ \hline \hline
\hfil 1116(2)   &\hfil 2.5(4)       & \hfil 4.6(8)      \\
\hfil 1529.7(7) &\hfil -            & \hfil 2.6(7)      \\
\hfil 1649.9(4) &\hfil -            & \hfil 7(1)        \\
\hfil 4110.8(3) &\hfil 8(1)         & \hfil8(2)        \\
\hfil 5439.6(4)$^d$ &\hfil -            &\hfil 4(2)        \\
\hfil 5711.6(9)$^e$ &\hfil -            &\hfil 3.7(12)$^c$ \\
\hfil 5770(1)   &\hfil -            & \hfil3.8(12)$^{c}$ \\
\hfil 5952.5(6)$^d$ &\hfil -            & \hfil6(2)$^{c}$    \\
\hfil 6018(2)   &\hfil 4.7(9)       &\hfil 6(2)$^c$    \\
\hfil 6067(2)   &\hfil -            & \hfil 3.9(13)$^c$ \\
\hfil 6220(2)   &\hfil -            & \hfil 2.6(10)$^c$ \\
\hfil 6463(3)   &\hfil -            & \hfil 2.0(9)$^c$  \\ \hline
\multicolumn{3}{l}{$^a$For I$_{abs}$ multiply by 0.049(5).}\\
\multicolumn{3}{l}{$^b$For I$_{abs}$ multiply by 0.043(5).}\\
\multicolumn{3}{l}{$^c$Intensity obtained from  $\beta$-gated spectrum.}\\
\multicolumn{3}{l}{$^d$Energy compatible with the escape peaks from the}\\
\multicolumn{3}{l}{\;\,tentative 6463-keV $\gamma$-ray.}\\
\multicolumn{3}{l}{$^e$Energy compatible with the single escape peak from}\\
\multicolumn{3}{l}{\;\,the tentative 6220-keV $\gamma$-ray.}
\\
\end{tabular}

\label{tab:Unassigned_gamma_rays}
\end{table}
\subsubsection{$\beta$-delayed neutron emission from $^{133}$In}
We have used the same procedure described above for $^{132}$In to determine the $\beta$-delayed neutron emission probabilities from the $^{133}$In (9/2$^{+}$) ground state and (1/2$^{-}$) isomer. The most intense $\gamma$ rays have been considered in the analysis. For $^{133}$Sn~$\rightarrow$~$^{133}$Sb decay, the absolute intensity of 12(2)$\%$ for the 962-keV transition was adopted from \cite{Blomqvist1983}, while for the $^{132}$Sn~$\rightarrow$~$^{132}$Sb an absolute intensity of 48.8(12)$\%$ for the 340-keV transition was taken from \cite{PhysRevC.39.1963}. The relative decay activity of the tin daughter nuclei was corrected for the tape movement. In particular, the supercycle structure of the proton beam has to be considered for the evaluation of the unobserved activity. Our analysis yields P$_n$=90(3)$\%$ for the decay of the $^{133}$In 9/2$^+$ g.s., and P$_n$=93(3)$\%$ for the decay of the (1/2$^-$) isomer. The results differ from those previously reported by us in \cite{Piersa2019}, which were obtained from the same dataset but where the supercycle structure was not fully taken into account. The re-evaluated results are in agreement with the P$_n$=85(10)$\%$ value in \cite{Hoff1996}.

\subsection{Lifetime measurements}
Lifetimes of excited levels in $^{132}$Sn have been investigated by means of the Advanced Time-Delayed $\beta\gamma\gamma$(t) fast-timing method \cite{Mach1989,Moszynski1989,Fraile2017}. The lifetimes were mainly obtained from the time differences between the fast $\beta$ and LaBr$_3$(Ce) detectors. A coincidence condition on the HPGe detectors is applied. The HPGe detectors do not participate in the timing information, but are essential in this complex level scheme due to their energy resolution to obtain the required $\gamma$-ray selectivity. 

\begin{figure}[ht!]
\includegraphics[width=\columnwidth]{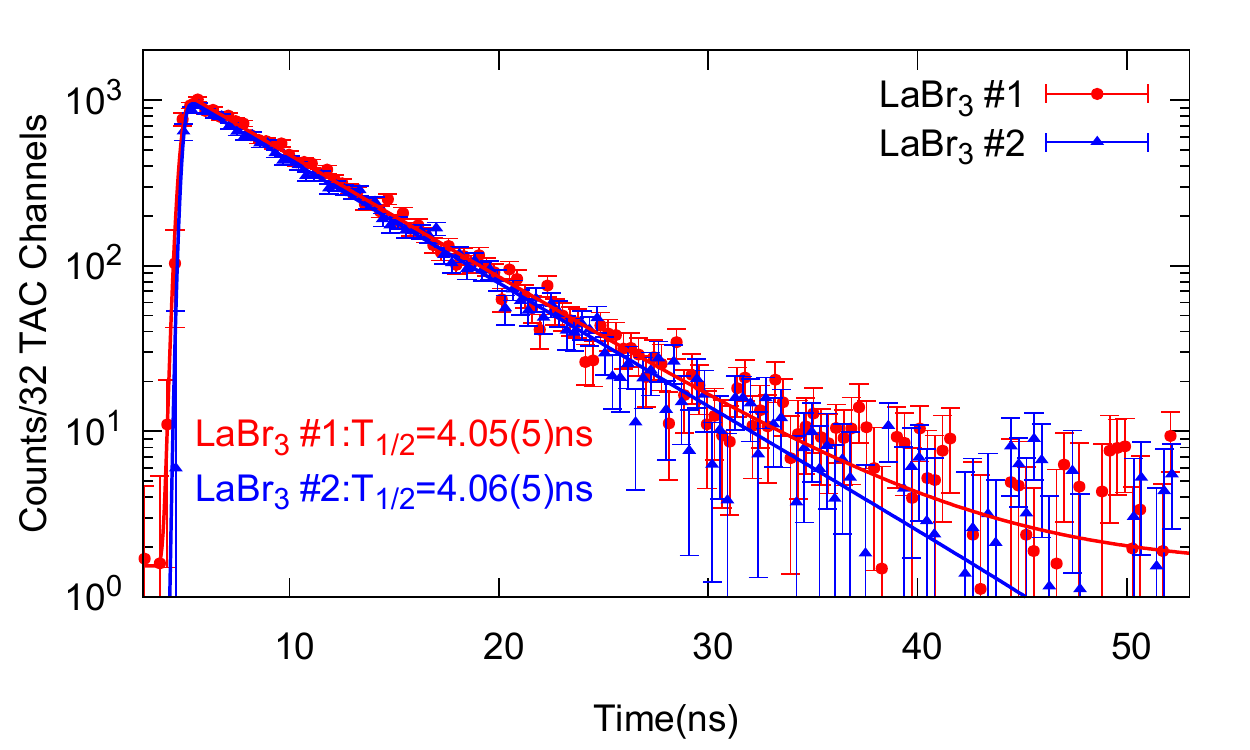}
\caption{\label{fig:4416_lifetime} Time delay spectra between the $\beta$ and each of the two LaBr$_3$(Ce) detectors for $\beta\gamma\gamma$(t) events. These distribution are built by selecting the events when the 526-keV transition in the HPGe detectors and the 375-keV transition in the LaBr$_3$(Ce) detector. The lifetime is obtained by a $\chi^2$ fit of the whole time distribution to a exponential decay convoluted with a Gaussian function plus a constant background to account for the random background.}
\end{figure}

\begin{figure}[ht!]

    \centering
    \includegraphics[width=\columnwidth]{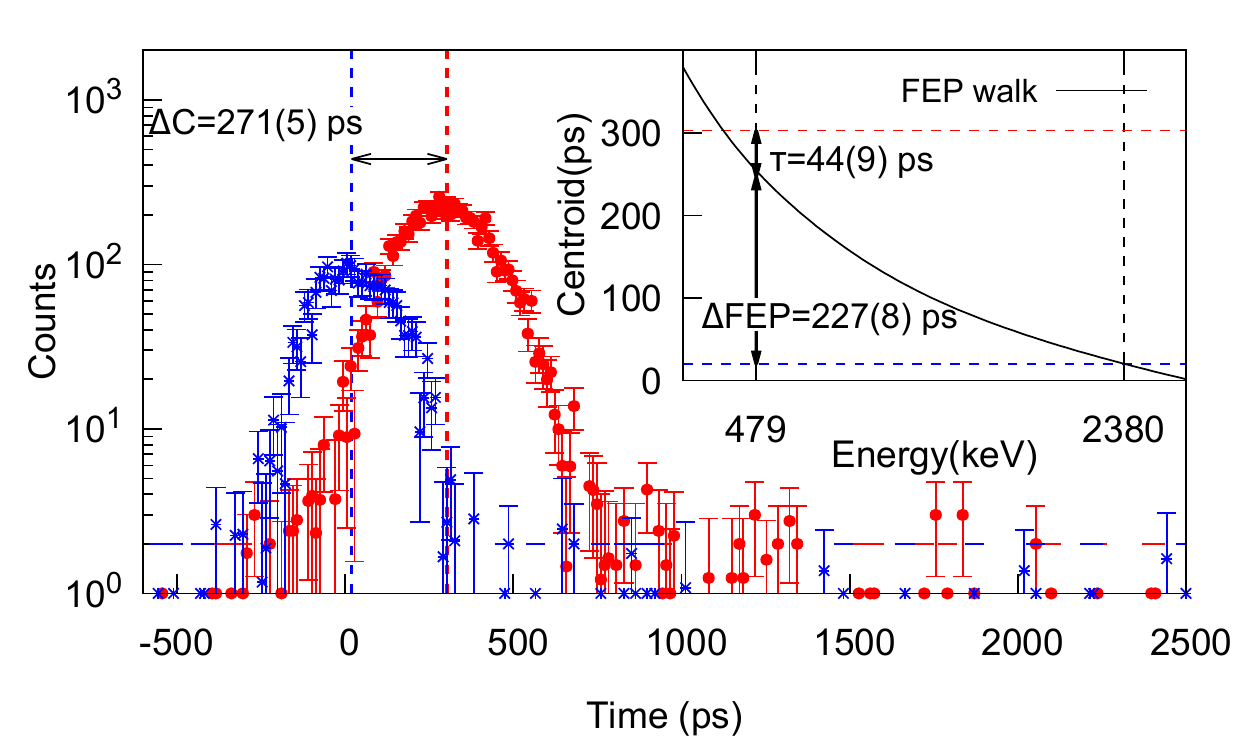}
    \caption{Time delayed $\beta\gamma\gamma$(t) spectra used to measure the lifetime of the 4830-keV 4$^-$ excited level. The blue spectra depicts the prompt time distribution used as reference, which is obtained by gating on the 479 and 2380-keV transitions in the HPGe and the LaBr$_3$(Ce) detectors respectively. The red one correspond to events obtained after reversing the gates, the 2380 keV one applied to the HPGe and the 479 keV one to the LaBr$_3$(Ce). The shift measured between the centroid position of each distribution $\Delta$C is caused by the lifetime of the level, but also due the time walk $\Delta$FEP between both energies. The lifetime $\tau$ is obtained by subtracting to the centroid shift the contribution due to the FEP time response.
    }
    \label{fig:4830_lifetime}
    
\end{figure}

The use of two different LaBr$_3$(Ce) detectors gives us the possibility to obtain two independent measurements for the same lifetime, one per $\beta$-LaBr$_3$(Ce) combination. In addition, $\gamma\gamma$(t) time differences between the two LaBr$_3$(Ce) detectors are used when possible. To illustrate the analysis and the results here obtained, we discuss the half-life of the 4416-keV 4$^+$ and 4831-keV 4$^-$ levels, depicted in Figure \ref{fig:4416_lifetime} and Figure \ref{fig:4830_lifetime} respectively.\par

To measure the 4416-keV 4$^+$ level half-life using $\beta\gamma\gamma$(t) events a time distribution was generated by selecting the 526- and 375-keV transitions in the HPGe and LaBr$_3$(Ce) detectors respectively. Corrections were included to account for the contribution of Compton background. The 4416-keV 4$^+$ state lifetime is free from the influence of other long-lives states and shows up as an exponential tail which can be fitted to measure the half-life. The analysis was done separately for each of the two LaBr$_3$(Ce) as well as for each of the two available data sets. In the case of $\gamma\gamma$(t) events, the 299-keV and 375-keV transitions are selected. Here no extra gate in the HPGe energies is needed thanks to the large peak to background ratio. The lifetime is measured from time difference distributions with the direct and reversed energy selection on the LaBr$_3$(Ce) detectors, giving two independent measurements for each of the two data sets. 
Using the two experimental data sets, our analysis yields 8 independent measurements for the half life, 4 from $\beta\gamma$(t) events and another 4 from $\gamma\gamma$(t), all of them consistent with each other. The final value is obtained from the weighted average from these measurements, yielding a final value of 3.99(2)~ns, which is in good agreement with the 3.95(13)~ns reported by Fogelberg \textit{et al.} \cite{Fogelberg1995}.\par

Shorter lifetimes were measured using the centroid shift method. In Figure \ref{fig:4830_lifetime}, the analysis to extract the lifetime of the 4830-keV 4$^-$ level is is illustrated for $\beta\gamma\gamma$(t) events for one of the LaBr$_3$(Ce) detectors. In this case the mean life is derived from the centroid shift of the time distribution with respect to a prompt cascade, corrected by the FEP time response calibration. 

The analysis has been repeated for both LaBr$_3$(Ce) detectors and both data sets, and also using the $\gamma\gamma$(t) method, yielding six independent values for the 4830-keV 4$^-$ level mean-life. The weighted average of T$_{1/2}$=27(2)~ps is adopted as the final result. The half-life is in very good agreement with the 26(5) ps reported by Fogelberg \textit{et al.} \cite{Fogelberg1995}. It should be noted that the uncertainties in \cite{Fogelberg1994} are too small to be compatible with a fast-timing measurement; the values within brackets in \cite{Fogelberg1995} are actually the errors in ps \cite{Priv_Comm_Mach}, so  the uncertainties from \cite{Fogelberg1995} are adopted.
\par

The lifetime of the 4848-keV 8$^+$ level is beyond the fast-timing time range, but it can be investigated using $\beta\gamma$(t) coincidences between the plastic scintillator and the HPGe detectors. Three $\beta$-HPGe(t) time differences spectra were obtained by selecting  the 132-, 299- and 375-keV $\gamma$ rays, respectively. The half-life was measured by fitting the delayed slope of the spectra in a long time range of $\sim$30~$\mu$s. The contribution of random coincidences to the time spectra in this range had to be carefully taken into account. Our analysis yields T$_{1/2}$=2.108(14)~$\mu$s for the 4848-keV level in agreement with the value of 2.080(17)~$\mu$s reported in the latest evaluation \cite{NDS132}.  \par

The analysis procedures were extended to the other observed levels, provided sufficient statistics were available. The data sets from the two experimental runs have been combined for the analysis. The lifetimes of 7 different states were measured. In addition, upper limits for additional 7 levels were obtained. The results obtained in this work for level lifetimes in $^{132}$Sn are compiled in Tables~\ref{tab:rates} and compared to the previous $\beta$-decay studies.

The overall agreement is quite good, with the exception of the 4919-keV 7$^+$ level whose half-life is 104(4) ps, much higher than the value of 62(7) ps reported earlier \cite{Fogelberg1994}, which was measured using BaF$_2$ scintillator detectors. Since this level can only be measured by analyzing the time distribution of the 203-keV $\gamma$-ray, where the contribution of Compton background is very important, the difference may stem from time corrections in this difficult energy range.\par

The reduced transition probabilities for de-exciting $\gamma$-transitions in $^{132}$Sn have been determined using the measured lifetimes, branching ratios and energies, and using theoretical internal conversion coefficients \cite{KIBEDI2008202}. The most likely spin-parity assignments are employed (see Section \ref{sec:discussion}). The transition rates are calculated assuming a pure multipolarity character of the transitions.

\LTcapwidth=\textwidth
\begin{longtable*}{ccccccccccc}
\\
\caption{\label{tab:rates} Half-lives and reduced transition probabilities of the transitions in $^{132}$Sn. The B(X$\lambda$)  values have been derived from the lifetimes and branching ratios obtained from $^{132}$In decay in this work \ref{tab:132sngammas}, as well as the theoretical internal conversion coefficients, calculated using Bricc \cite{KIBEDI2008202}. Transition rates have been calculated assuming a pure multipolarity character of the transitions, using the assignments from \cite{Fogelberg1994}. For those levels where no previous assignment had been made, the B(X$\lambda$) values corresponding to the most likely multipolarities are presented.}
\\
\\\hline
E$_i$(keV) &     Config$_i$                           &J$^{\Pi}_i$ & T$_{1/2}$ & T$_{1/2}$ (literature) & E$_f$(keV) &Config$_f$ & J$^{\Pi}_f$ & E$_{\gamma}$(keV) & X$\lambda$ & B(X$\lambda$) (W.u.)\\\hline \hline
\endfirsthead
\caption{(Continued)}\\
\\\hline
E$_i$(keV) &     Config$_i$                           &J$^{\Pi}_i$ & T$_{1/2}$ & T$_{1/2}$ (literature) & E$_f$(keV) &Config$_f$ & J$^{\Pi}_f$ & E$_{\gamma}$(keV) & X$\lambda$ & B(X$\lambda$) (W.u.)\\\hline \hline
\endhead

  \hline \multicolumn{11}{c}{\textit{Continued on next page}} \\
  \endfoot
\hline          
\multicolumn{11}{l}{$^a$ The uncertainties are taken from \cite{Fogelberg1995} since those in \cite{Fogelberg1994} contain several typographical errors \cite{Priv_Comm_Mach}.} \\    
\multicolumn{11}{l}{$^b$ Calculated from the B(E3) rate measured in Coulomb excitation \cite{Rosiak2018}} \\  
\multicolumn{11}{l}{$^c$ Assigned multipolarity in \cite{Fogelberg1994}} \\  
  \hline   
  \endlastfoot
\hline

4351.6 & Octupole                   & 3$^-$   & \textless{}5 ps  & \textless{}5 ps \cite{Fogelberg1994}       & 0      & g.s.                       & 0$^+$  & 4351.5 & E3$^c$ & \textgreater{}7.1               \\
       &  vibration                 &         &                  &3.4($^{+20}_{-9}$) ps \cite{Rosiak2018}$^b$ & 4041.6 & $\nu f_{7/2}h^{-1}_{11/2}$ & 2$^+$  & 310.5  & E1$^c$ & \textgreater 1.2$\cdot$10$^{-4}$ \\\nopagebreak \hline
4416.6 & $\nu f_{7/2}h^{-1}_{11/2}$ & 4$^+$   & 3.99(2) ns       & 3.95(13) ns \cite{Fogelberg1994}           & 0      & g.s.                       & 0$^+$  & 4416.7 & E4     & 7.7(4)                      \\\nopagebreak
       &                            &         &                  &                                            & 4041.6 & $\nu f_{7/2}h^{-1}_{11/2}$ & 2$^+$  & 374.9  & E2$^c$ & 0.40(2)                 \\\nopagebreak
       &                            &         &                  &                                            & 4351.6 & Octupole vibration         & 3$^-$  & 64.4   & E1     & 2.57(13)$\cdot$10$^{-6}$    \\ \hline
4715.9 & $\nu f_{7/2}h^{-1}_{11/2}$ & 6$^+$   & 21.3(4) ns       & 20.1(5) ns \cite{Fogelberg1994}            & 4416.6 & $\nu f_{7/2}h^{-1}_{11/2}$ & 4$^+$  & 299.3  & E2$^c$ & 0.268(6)               \\ \hline
4830.5 & $\nu f_{7/2}d^{-1}_{3/2}$ & 4$^-$   & 27(2) ps         & 26(5) ps \cite{Fogelberg1994}$^a$          & 4351.6 & Octupole vibration         & 3$^-$  & 478.9  & M1$^c$ & 7.3(5)$\cdot$10$^{-3}$        \\\nopagebreak
       &                            &         &                  &                                            & 4416.6 & $\nu f_{7/2}h^{-1}_{11/2}$ & 4$^+$  & 414.5  & E1     & 2.3(3)$\cdot$10$^{-6}$     \\ \hline
4848.3 & $\nu f_{7/2}h^{-1}_{11/2}$ & 8$^+$   & 2.108(14) $\mu$s & 2.080(17) $\mu$s \cite{NDS132}        & 4715.9 & $\nu f_{7/2}h^{-1}_{11/2}$ & 6$^+$  & 132.4  & E2$^c$ & 0.1039(14) \\ \hline
4885.7 & $\nu f_{7/2}h^{-1}_{11/2}$ & 5$^+$   & \textless{}30 ps & \textless{}40 ps \cite{Fogelberg1994}      & 4416.6 & $\nu f_{7/2}h^{-1}_{11/2}$ & 4$^+$  & 469.1  & M1     & \textgreater{}6.5$\cdot$10$^{-3}$     \\\nopagebreak
       &                            &         &                  &                                            &        &                            &        &        & E2     & \textgreater{}19                   \\\nopagebreak
       &                            &         &                  &                                            & 4715.9 & $\nu f_{7/2}h^{-1}_{11/2}$ & 6$^+$  & 169.5  & M1     & \textgreater{}4.6$\cdot$10$^{-3}$                \\\nopagebreak
       &                            &         &                  &                                            &        &                            &        &        & E2     & \textgreater{}94               \\ \hline
4918.8 & $\nu f_{7/2}h^{-1}_{11/2}$ & 7$^+$   & 104(4) ps        & 62(7) ps \cite{Fogelberg1994}$^a$          & 4715.9 & $\nu f_{7/2}h^{-1}_{11/2}$ & 6$^+$  & 202.9  & M1     & 1.74(9)$\cdot$10$^{-2}$                      \\\nopagebreak
       &                            &         &                  &                                            & 4848.3 & $\nu f_{7/2}h^{-1}_{11/2}$ & 8$^+$  & 70.9   & M1     & 6.0(7)$\cdot$10$^{-2}$                     \\ \hline
4942.4 & $\nu f_{7/2}d^{-1}_{3/2}$ & 5$^-$   & 23(2) ps         & 17(5) ps \cite{Fogelberg1994}$^a$          & 4351.6 & Octupole vibration         & 3$^-$  & 590.4  & E2     & 0.24(3)                       \\\nopagebreak
       &                            &         &                  &                                            & 4416.6 & $\nu f_{7/2}h^{-1}_{11/2}$ & 4$^+$  & 525.9  & E1$^c$ & 6.7(7)$\cdot$10$^{-5}$            \\\nopagebreak
       &                            &         &                  &                                            & 4715.9 & $\nu f_{7/2}h^{-1}_{11/2}$ & 6$^+$  & 226.5  & E1     & 1.7(2)$\cdot$10$^{-5}$            \\\nopagebreak
       &                            &         &                  &                                            & 4830.5 & $\nu p_{3/2}h^{-1}_{11/2}$ & 4$^-$  & 111.3  & M1     & 5.2(7)$\cdot$10$^{-2}$  \\ \hline
5387.3 & ($\nu f_{7/2}s^{-1}_{1/2}$)& (4$^-$) & \textless{}17 ps &                                            & 4351.6 & Octupole vibration         & 3$^-$  & 1036.0 & M1     & \textgreater{}0.8$\cdot$10$^{-3}$              \\\nopagebreak
       &                            &         &                  &                                            &        &                            &        &        & E2     & \textgreater{}0.5                 \\\nopagebreak
       &                            &         &                  &                                            & 4830.5 & $\nu f_{7/2}d^{-1}_{3/2}$  & 4$^-$  & 557.1  & M1     & \textgreater{}4.5$\cdot$10$^{-4}$              \\\nopagebreak
       &                            &         &                  &                                            &        &                            &        &        & E2     & \textgreater{}0.9                 \\\nopagebreak
       &                            &         &                  &                                            & 4942.4 & $\nu f_{7/2}d^{-1}_{3/2}$  & 5$^-$  & 444.6  & M1     & \textgreater{}1.5$\cdot$10$^{-3}$               \\\nopagebreak
       &                            &         &                  &                                            &        &                            &        &        & E2     & \textgreater{}5.1                  \\\nopagebreak
       &                            &         &                  &                                            & 4949.0 & $\nu f_{7/2}d^{-1}_{3/2}$  & (3$^-$)& 437.2  & M1     & \textgreater{}1.7$\cdot$10$^{-3}$              \\\nopagebreak
       &                            &         &                  &                                            &        &                            &        &        & E2     & \textgreater{}5.6                \\ \hline 
       
       5398.9 & ($\pi g_{7/2}g^{-1}_{9/2}$) & (6$^+$) & \textless{}17 ps &                                    & 4715.9 & $\nu f_{7/2}h^{-1}_{11/2}$  & 6$^+$   & 683.0  & M1 & \textgreater{}4.0$\cdot$10$^{-3}$                \\\nopagebreak
       &                             &         &                  &                                    &        &                             &         &        & E2 & \textgreater{}5.6                  \\ \hline
5478.4 & ($\pi g_{7/2}g^{-1}_{9/2}$) & (8$^+$) & \textless{}14 ps &                                    & 4848.3 & $\nu f_{7/2}h^{-1}_{11/2}$  & 8$^+$   & 630.2  & M1 & \textgreater{}6.4$\cdot$10$^{-3}$               \\
       &                             &         &                  &                                    &        &                             &         &        & E2 & \textgreater{}10              \\ \hline
5628.9 & ($\pi g_{7/2}g^{-1}_{9/2}$) & (7$^+$) & 9(3) ps          & 13(4) ps \cite{Fogelberg1994}$^a$  & 4715.9 & $\nu f_{7/2}h^{-1}_{11/2}$  & 6$^+$   & 913.1  & M1 & 2.0($^{+9}_{-5}$)$\cdot$10$^{-3}$        \\*  \nopagebreak 
      &                             &         &                  &                                    &        &                             &         &        & E2 & 1.5($^{+7}_{-4}$)                        \\*  \nopagebreak 
      &                             &         &                  &                                    & 4848.3 & $\nu f_{7/2}h^{-1}_{11/2}$  & 8$^+$   & 780.6  & M1 & 1.3($^{+6}_{-3}$)$\cdot$10$^{-3}$                \\*  \nopagebreak
       &                             &         &                  &                                    &        &                             &         &        & E2 & 1.4($^{+6}_{-3}$)                            \\*  \nopagebreak 
      &                             &         &                  &                                    & 4918.8 & $\nu f_{7/2}h^{-1}_{11/2}$  & 7$^+$   & 710.1  & M1 & 7($^{+3}_{-2}$)$\cdot$10$^{-4}$               \\*  \nopagebreak
      &                             &         &                  &                                    &        &                             &         &        & E2 & 0.9($^{+4}_{-2}$)                          \\\nopagebreak 
      &                             &         &                  &                                    & 5398.9 & ($\pi g_{7/2}g^{-1}_{9/2}$) & (6$^+$) & 229.8  & M1 & 6($^{+3}_{-2}$)$\cdot$10$^{-3}$          \\*  \nopagebreak 
     &                             &         &                  &                                    &        &                             &         &        & E2 & 76($^{+34}_{-19}$)                            \\\nopagebreak  
     &                             &         &                  &                                    & 5478.4 & ($\pi g_{7/2}g^{-1}_{9/2}$) & (8$^+$) & 150.3  & M1 & 9($^{+4}_{-2}$)$\cdot$10$^{-3}$         \\\nopagebreak 
       &                             &         &                  &                                    &        &                             &         &        & E2 & 215($^{+95}_{-54}$)                \\\hline
5753.9 & ($\nu p_{3/2}h^{-1}_{11/2}$)& (6$^+$) & \textless{}20 ps &                                    & 4715.9 & $\nu f_{7/2}h^{-1}_{11/2}$  & 6$^+$   & 1038.2 & M1 & \textgreater{}7.1$\cdot$10$^{-4}$               \\\nopagebreak
       &                             &         &                  &                                    &        &                             &         &        & E2 & \textgreater{}0.4                 \\\nopagebreak
       &                             &         &                  &                                    & 5398.9 & ($\pi g_{7/2}g^{-1}_{9/2}$) & (6$^+$) & 354.3  & M1 & \textgreater{}5.5$\cdot$10$^{-3}$              \\\nopagebreak
       &                             &         &                  &                                    &        &                             &         &        & E2 & \textgreater{}28                 \\  \hline     
6235.5 &                             & (7$^+$) & \textless{}10 ps &                                    & 4715.9 & $\nu f_{7/2}h^{-1}_{11/2}$  & 6$^+$   & 1519.6 & M1 & \textgreater{}1.6$\cdot$10$^{-4}$          \\\nopagebreak
       &                             &         &                  &                                    &        &                             &         &        & E2 & \textgreater{}0.04                       \\\nopagebreak
       &                             &         &                  &                                    & 4918.8 & $\nu f_{7/2}h^{-1}_{11/2}$  & 7$^+$   & 1317.1 & M1 & \textgreater{}1.4$\cdot$10$^{-4}$                    \\\nopagebreak
       &                             &         &                  &                                    &        &                             &         &        & E2 & \textgreater{}0.05                          \\\nopagebreak
       &                             &         &                  &                                    & 5398.9 & ($\pi g_{7/2}g^{-1}_{9/2}$) & (6$^+$) & 836.3  & M1 & \textgreater{}1.1$\cdot$10$^{-3}$              \\\nopagebreak
       &                             &         &                  &                                    &        &                             &         &        & E2 & \textgreater{}1.0                          \\\nopagebreak
       &                             &         &                  &                                    & 5753.9 & ($\nu p_{3/2}h^{-1}_{11/2}$)& (6$^+$) & 481.8  & M1 & \textgreater{}4.8$\cdot$10$^{-3}$              \\\nopagebreak
       &                             &         &                  &                                    &        &                             &         &        & E2 & \textgreater{}13                  \\ \hline
6709.7 &                             & (7$^-$) & \textless{}13 ps &                                    &4918.8  &$\nu f_{7/2}h^{-1}_{11/2}$   & 7$^+$   & 1791.0 & E1 & \textgreater{}3.6$\cdot$10$^{-7}$           \\\nopagebreak
        &                            &         &                 &                                     & 4942.4 & $\nu f_{7/2}d^{-1}_{3/2}$   & 5$^-$   & 1767.2 & E2 & \textgreater{}0.06                      \\\nopagebreak

\end{longtable*}

\section{Discussion}
\label{sec:discussion}
\begin{figure*}[ht!]
    \includegraphics[width=\textwidth]{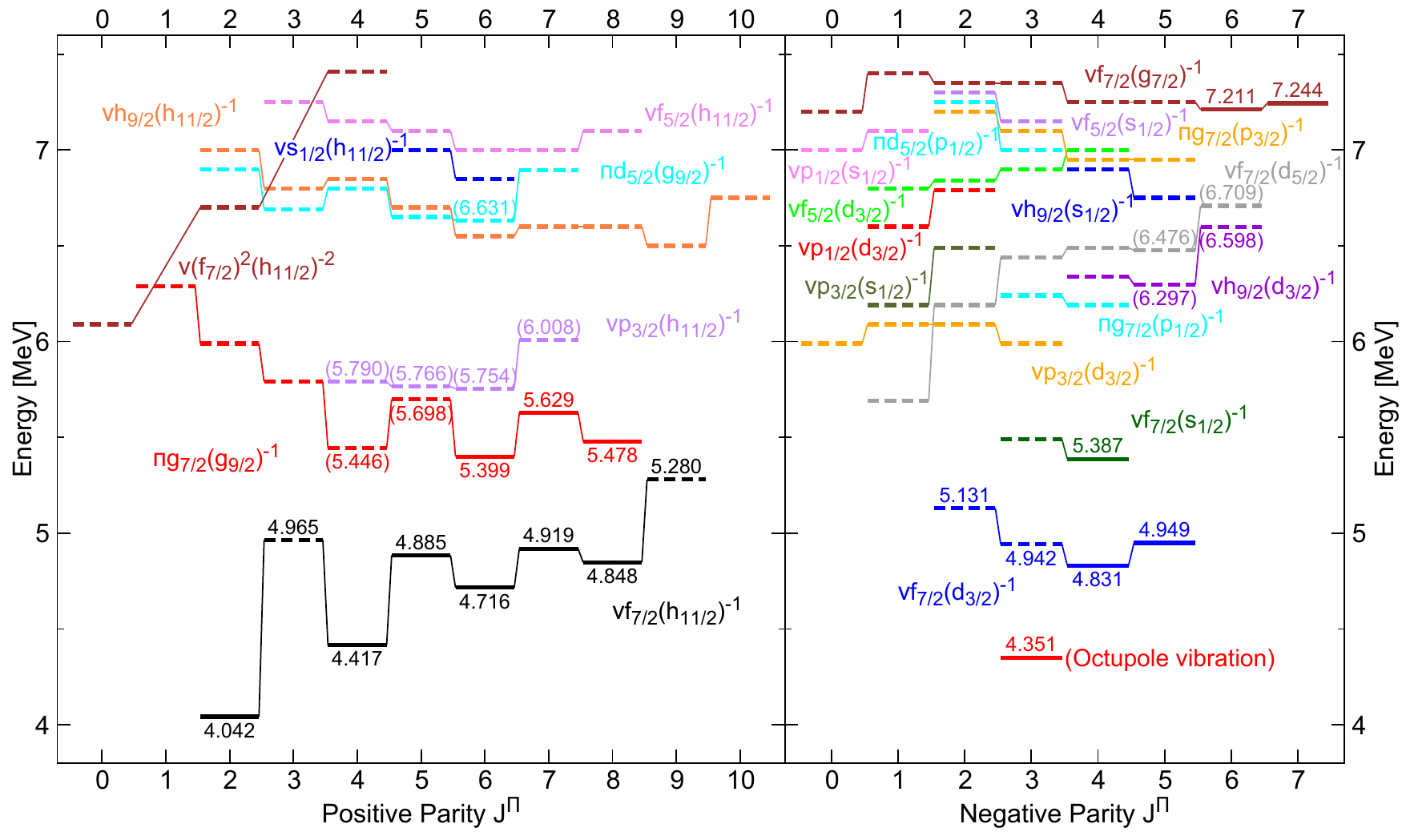}
    \caption{Calculated energies for the different particle-hole multiplet states in $^{132}$Sn, adopted from J. Blomqvist \cite{Priv_Comm_Blomqvist_132,MACH1995c179}. The energies and energy splitting within a given multiplet are estimated by scaling the analogous particle-hole states in $^{208}$Pb and taking into account single particle energies from neighboring nuclei. Previously identified levels are plotted with continuous lines. The experimental energies from our work are shown for the newly identified states. The levels whose energies appear between brackets correspond to tentative assignments.}
    \label{fig:Blomqvist}
\end{figure*}

All new levels in $^{132}$Sn observed in this investigation are candidates for the remaining unidentified states within the particle-hole multiplets. In Figure \ref{fig:Blomqvist} the energies for the 24 particle-hole multiplets in $^{132}$Sn expected to appear below the neutron separation energy are represented. The energies and the splitting of the different levels for the same multiplet are estimated by taking into account the single particle energies from neighboring nuclei, and the analogous particle-hole states in $^{208}$Pb, taken from \cite{MACH1995c179,Fogelberg1995,Priv_Comm_Blomqvist_132}, where a A$^{-1/3}$ scale is introduced to take into account the nuclear potential depth. These empirical calculations provide guidance for the location of the p-h states, which allows to propose spin-parity assignments to the new levels found in this work.

The other piece of information is provided by the vast number of new transitions which connect the new states to known levels. By assuming that the transitions are predominantly of dipole character (mainly of M1 multipolarity) and using the electromagnetic selection rules it is possible to make tentative spin-parity assignments of the newly-identified levels. The transition rates obtained from the measured lifetimes and lifetime limits of the new levels also provide constraints. 
Together with the information from the systematics of p-h states tentative configurations for the new levels are proposed. In the level schemes presented above, the tentative spin and parity assignments are already shown.\par

\subsection{Low energy neutron particle-hole states}

The unique identification of every observed state in $^{132}$Sn with a specific state from a given p-h multiplet is a very complex task. In previous $\beta$-decay studies \cite{Fogelberg1994,Bjornstad1986} the tentative assignment of several states was done by considering the analogies between the $^{132}$In [($\pi$g$_{9/2}$)$^{-1}\nu$f$_{7/2}$] decay and its two neighbors $^{131}$In [$\pi g^{-1}_{9/2}$] and $^{133}$Sn [$\nu f_{7/2}$] decays. 
The 6$^-$ and 7$^-$ levels observed above 7 MeV, attributed to the $\nu f_{7/2}g^{-1}_{7/2}$ configuration, receive the main fraction of the total $\beta$-feeding intensity. This strong transition would be equivalent to the Gamow-Teller transition in the $^{131}$In decay $\pi g^{-1}_{9/2}\rightarrow \nu g^{-1}_{7/2}$ \cite{Fogelberg2004}. Keeping up with the $^{131}$In decay analogy, the second strongest decay branch corresponds to the first-forbidden transition $\pi g^{-1}_{9/2}\rightarrow \nu h^{-1}_{11/2}$. In the $^{132}$In decay the equivalent transition would populate the $\nu f_{7/2}h^{-1}_{11/2}$ multiplet. Consequently the positive parity states that appear below 5 MeV are suggested to be members of this configuration. On the other hand, the strong 2435-keV $\gamma$ transition in $^{131}$Sn that connects the single hole states $\nu g^{-1}_{7/2}\rightarrow$ $\nu d^{-1}_{3/2}$ would be analogue to the 2380-keV and 2268-keV transitions in $^{132}$Sn, which indicates that the 4$^-$ and 5$^-$ found below 5 MeV are indeed members of the $\nu f_{7/2}d^{-1}_{3/2}$ p-h multiplet. \par

In this way there are two neutron p-h multiplets identified within this energy range. However, several of the expected levels from those configurations lack experimental identification. Specifically the 3$^+$ and 8$^+$ levels from the $\nu f_{7/2}h^{-1}_{11/2}$ coupling, as well as the 2$^-$ and 3$^-$ levels from the $\nu f_{7/2}d^{-1}_{3/2}$ multiplet, which are expected to be found at excitation energies around 5 MeV, remain unseen. In a previous investigation from the $^{248}$Cm spontaneous fission \cite{Bhattacharyya2001}, a level at 5280 keV was suggested to be the missing $\nu f_{7/2}h^{-1}_{11/2}$ 8$^+$ level. On the contrary, in our work we observe the population of this level in the $^{132}$In decay, which de-excites by a $\gamma$ ray to the 4848-keV state. This level has a very large apparent log$ft$ value, around 7.2, as it would be expected for a high spin such as 9$^+$, which suggests its identification as the missing 9$^+$ level. \par

The remaining three levels found within that energy range are the 4949-, 4965- and 5131-keV states. They are directly populated in the $\beta$-n decay of $^{133g,m}$In, but not directly fed from the $^{132}$In (7$^{-}$) decay. The 4949-keV level is observed in the $^{132}$In decay, but indirectly populated by a $\gamma$-transition from the 5387 keV (4$^-$) level without any noticeable direct $\beta$ feeding. 
The $\beta$-n feeding from the (1/2$^-$) and (9/2$^+$) $^{133}$In states points towards a low spin value for these three levels. Specifically the 5131-keV state is populated more favorably in the $\beta$-n decay of the (1/2$^-$) isomer, which, along with the $\gamma$ transition which connects that level with the g.s., suggests a lower spin for that level than for the other ones. This is an indication that the 5131-keV state is very likely the 2$^-$ member of the $\nu f_{7/2}d^{-1}_{3/2}$ multiplet. The systematics of the transitions that de-excite the 4949- and 4965-keV levels suggests angular momenta \textit{J}=3. Therefore one of them would correspond to the 3$^+$ state from $\nu f_{7/2}h^{-1}_{11/2}$ and the other matches the 3$^-$ from $\nu$f$_{7/2}$(p$_{3/2}$)$^{-1}$. There is no direct information to determine their parity, since both of them are linked to positive and negative parity levels. However, there are analogies between the 4949 keV and the 5$^-$ 4942-keV levels. Firstly, both are being populated from the 5387-keV state, interpreted as $\nu f_{7/2}s^{-1}_{1/2}$ (4$^-$), with similar intensities. Secondly both levels have a transition to the 4352-keV 3$^-$ collective state, and another one to the $\nu f_{7/2}d^{-1}_{3/2}$ 4$^-$ level whose intensity ratio is about 10:1 in both cases. Those similarities strongly suggest that 4949-keV level is indeed the 3$^-$ member from the $\nu f_{7/2}d^{-1}_{3/2}$ multiplet, and therefore the 4965 keV can only be the 3$^+$ member of the $\nu f_{7/2}h^{-1}_{11/2}$ p-h configuration.

The new assignments are shown in Figure \ref{fig:Blomqvist}.

\subsection{Particle-hole states from 5 to 6 MeV}

In the investigation by Fogelberg \textit{et al.} \cite{Fogelberg1995,Fogelberg1994}, the 5399-keV (6$^+$), 5478-keV (8$^+$) and 5629-keV (7$^+$) levels were identified as members of the proton p-h $\pi g_{7/2}g^{-1}_{9/2}$ multiplet. This assignment was supported by the intense feeding of the 5629 keV (7$^+$) state, which would be the equivalent to the strongest transition in the $^{133}$Sn decay, $\nu$f$_{7/2}\rightarrow\pi$g$_{7/2}$. Within this region we have found 7 new levels with expected angular momenta from 4 to 8. In addition to the proton $\pi g_{7/2}g^{-1}_{9/2}$ p-h levels, states from the neutron $\nu p_{3/2}h^{-1}_{11/2}$ and $\nu f_{7/2}s^{-1}_{1/2}$ configurations are also expected. All the levels found from 5.3 to 6 MeV could be related to p-h levels from those multiplets. There is not much information apart from $\gamma$-ray intensities and the lifetimes and lifetime limits that point towards dominant M1 transitions.  This, together with the possibility that the configurations are mixed, hinders a clearer assignment. 
Nevertheless several conclusions about them can be drawn. 

For the 5431-keV level, observed only in the $^{133}$In decay, a \textit{J} = 3 spin assignment can be assumed based on the $\gamma$ transitions systematics. Thus, this level can be either identified with the 3$^+$ state of the $\pi g_{7/2}g^{-1}_{9/2}$ or with the 3$^-$ of the $\nu f_{7/2}s^{-1}_{1/2}$ configuration.\par

At similar excitation we expect two 4$^+$ states belonging to the two positive parity multiplets mentioned above. In our analysis two levels have been found within this energy range with a tentative angular momenta \textit{J} = 4. There is a level at 5446 keV (4$^+$), populated indirectly by $\gamma$ transitions in $^{132}$In and directly in the $\beta$-n decay of $^{133g,m}$In. This level is a good candidate to be a member of the proton  $\pi g_{7/2}g^{-1}_{9/2}$ multiplet, but it can also be interpreted as the 4$^+$ level from  the neutron $\nu p_{3/2}h^{-1}_{11/2}$ configuration. At 5790 keV another level was identified, this one can be only observed in the $^{133}$In $\beta$-n decay, and its de-exciting transitions point towards a spin of (3,4). Therefore, this level can be identified either as the 4$^+$ from the $\nu p_{3/2}h^{-1}_{11/2}$ particle-hole coupling expected in this region, or the 3$^+$ from $\pi g_{7/2}g^{-1}_{9/2}$.

The (5$^+$) levels at 5698 keV and 5766 keV, which are observed in this work in both the $^{132}$In and $^{133}$In decays, can be related to the 5$^+$ states from the $\pi g_{7/2}g^{-1}_{9/2}$ and the $\nu p_{3/2}h^{-1}_{11/2}$ multiplets. The two remaining levels in this region are the 5754-keV (6$^+$) and the 6008-keV (7$^+$) ones, which we tentatively suggest as members of the $\nu p_{3/2}h^{-1}_{11/2}$ particle-hole configuration. 

These tentative assignments are reflected in Figure \ref{fig:Blomqvist}.

\subsection{States from 6 to 7 MeV}
Moving up to the next energy interval the identification becomes even more complicated, due to all the possible p-h multiplets that are expected in this region, and the likely admixture of configurations.
Nevertheless, since most of these levels are populated only in the direct $\beta$-decay of $^{132}$In, they are constrained by the selective nature of $\beta$ decay that favors, in this case, states with a large spin (6-8). Because there are not so many p-h multiplets at this energy that could give rise to levels with such a large spin (Figure \ref{fig:Blomqvist}) we can draw some conclusions about them. \par

Regarding the negative parity states, there are two high-lying levels with most likely (6$^{-}$) assignments observed only in the $^{132}$In decay, at 6598 and 6709 keV. Only two neutron multiplets, the $\nu h_{9/2}d^{-1}_{3/2}$ and $\nu f_{7/2}d^{-1}_{5/2}$ have a negative parity member with \textit{J} = 6, and therefore these two levels can only be related to those. The remaining two negative parity levels appearing at  6297 and 6476 keV are present in both the $^{132}$In and $^{133}$In $\beta$ decays. The systematics of the transitions that de-excite and populate them suggest a (5$^-$) spin-parity and therefore they are likely to be members of the $\nu h_{9/2}d^{-1}_{3/2}$ and $\nu f_{7/2}d^{-1}_{5/2}$ multiplets as well.\par

On the positive parity side the situation is more involved due to the larger amount of multiplets predicted at these high energies. There are 8 different states with an assumed positive parity. These were only observed in the direct $\beta$ decay of $^{132}$In, out of which 5 have been identified in this work for the first time. Relying on the systematics of the transitions between these levels, along with the selectivity of the $\beta$-decay population from the (7$^-$) g.s. in $^{132}$In, they can have angular momenta from 5 to 8. Four different positive parity p-h multiplets are predicted in this energy range, three of them expected to arise from neutron configurations ($\nu h_{9/2}h^{-1}_{11/2}$, $\nu f_{5/2}h^{-1}_{11/2}$ and $\nu p_{3/2}h^{-1}_{11/2}$), and another one coming from the proton $\pi d_{5/2}g^{-1}_{9/2}$ coupling. Almost all of them can be give rise to (5,6,8)$^+$ states, which hampers their identification. Moreover the states are probably mixed.\par

The only noticeable difference observed among the positive parity levels is the larger feeding measured for the 6631-keV state, around 1$\%$ with a log$ft$ of the order of 6.4 (Figure \ref{fig:LvSc132In}). This enhanced population can be interpreted on the grounds of the equivalence mentioned before between the $^{132}$In and $^{133}$Sn decays. In $^{133}$Sn, the $\beta$ decay is dominated by the first forbidden $\nu$f$_{7/2}\rightarrow\pi$g$_{7/2}$ transition populating the g.s. in $^{133}$Sb with 86$\%$ of the total $\beta$ intensity and log$ft$ $\approx$ 5.5 \cite{Sanchez-Vega1999}, which is similar to the population of the $\pi g_{7/2}g^{-1}_{9/2}$ multiplet in the $^{132}$In decay. Moreover, the second strongest transition found in $^{133}$Sn decay corresponds to the $\nu f_{7/2}\rightarrow \pi d_{5/2}$ transition, receiving 11$\%$ of the total feeding with log$ft$ $\approx$ 6.1 \cite{Sanchez-Vega1999}. This transition would be equivalent to the population of the proton $\pi d_{5/2}g^{-1}_{9/2}$ configuration in the $^{132}$In $\beta$-decay. Comparing the population of the 6631-keV level with the population of 5629-keV level, identified as the (7$^+$) state of the $\pi g_{7/2}g^{-1}_{9/2}$ multiplet, we can see that they keep the 8:1 $\beta$ intensity ratio, and similar log$ft$ values for the level at 5629 keV. Such a large population might be an indication of a predominant $\pi d_{5/2}g^{-1}_{9/2}$ configuration for the 6631 keV (6$^+$) state. \par

\section{Conclusions}
\label{sec:conclusions}
Experimental information about the $^{132}$Sn structure plays a crucial role in the shell-model interpretation of nuclei around \textit{N} = 82, because it provides direct knowledge about the particle-hole couplings for both protons and neutrons. In this work the properties of excited states in \textsuperscript{132}Sn have been studied from the $\beta$ decay of \textsuperscript{132}In. By taking advantage of the isomer selectivity capabilities of the ISOLDE RILIS, independent investigations of the $\beta$-decay of the \textsuperscript{133}In$^g$ (9/2$^+$) g.s. and the \textsuperscript{133m}In (1/2$^-$) isomer were performed as well. Thanks to both decay modes, the knowledge of the \textsuperscript{132}Sn structure has been largely expanded in this work. A total of 17 new levels and 68 new $\gamma$-transitions have been added (including those already quoted in the previous publication \cite{Piersa2019} derived from the same experiment). A complete fast-timing investigation of the excited levels in $^{132}$Sn has been performed as well, confirming and extending previous results. \par

An interpretation of the level structure is provided in terms of particle-hole configurations arising from core breaking states both from the \textit{N} = 82 and \textit{Z} = 50 shells across the gap. The interpretation is based on empirical calculations \cite{MACH1995c179,Fogelberg1995,Priv_Comm_Blomqvist_132} leading to positive and negative parity particle-hole multiplets, where the energies are obtained from the single-particle energies from neighboring nuclei and from the analogous particle-hole states in $^{208}$Pb. These empirical calculations, together with the experimental information on the $\beta$ and $\beta$-n feeding, the level lifetimes and the $\gamma$ decay branches provide guidance for the identification of the levels as proton-hole members of the multiplets, facilitating tentative spin-parity assignments to the new levels. 

The number of states is consistent with the one calculated from the angular momentum couplings. The identification of all remaining missing levels from the $\nu f_{7/2}h^{-1}_{11/2}$ and $\nu f_{7/2}d^{-1}_{3/2}$ neutron-hole configurations has been completed. 
A tentative interpretation has been provided for the rest of observed states, where we have been able to observe most of the expected p-h levels with angular momenta close to 7. 

Most of the missing information is related to the anticipated low spin states, which are very unlikely to be populated in the $^{132}$In (7$^-$) $\beta$-decay. However, the identification of many $\gamma$-rays around 7~MeV from the $^{133g,m}$In $\beta$-decay strongly suggests feeding of those missing low-spin multiplet states. Enhanced statistics for this decay will be beneficial to provide more firm assignments. \par

In conclusion the knowledge about states in the doubly magic \textsuperscript{132}Sn has been largely expanded by the investigation of the $\beta$-decay of \textsuperscript{132}In and $\beta$-n decay of \textsuperscript{133}In performed at the ISOLDE facility at CERN. The identification of particle-hole multiplets both for protons and neutrons and the transition rates connecting different p-h configurations and states within multiplets may provide input on the two-body matrix elements and single particle states. The new data challenges the theoretical description of \textsuperscript{132}Sn, which is relevant for the understanding of nuclear structure in the region.

\section{Acknowledgments}
We acknowledge the support of the ISOLDE Collaboration and the ISOLDE technical teams, and by the European Union Horizon 2020 research and innovation programme under grant agreement No 654002. This work was partially funded by the Spanish government via FPA2015-65035-P, FPA-64969-P, FPA2017-87568-P and RTI2018-098868-B-I00 projects, the Polish National Science Center under Contracts No. UMO-2015/18/E/ST2/00217, UMO-2015/18/M/ST2/00523 and UMO-2019/33/N/ST2/03023, the Portuguese FCT via CERN/FIS-NUC/0004/2015 project, the German BMBF under contract 05P18PKCIA, the Romanian IFA Grant CERN/ISOLDE and by grants from the U.K. Science and Technology Facilities Council, the Research Foundation Flanders (FWO, Belgium), the Excellence of Science program (EOS, FWO-FNRS, Belgium), and the GOA/2015/010 (BOF KU Leuven). J.B. acknowledges support by the Universidad Complutense de Madrid under the predoctoral grant CT27/16-CT28/16.


%

\end{document}